# Thermodynamic model for electron emission, negative and positive ion formation in keV molecular collisions

Z. Juhász

*Institute for Nuclear Research, Hungarian Academy of Sciences (MTA Atomki), P.O. Box 51, H-4001 Debrecen, Hungary*

**Abstract**:
A statistical-type model is developed to describe the ion production and electron emission in collisions of (molecular) ions with atoms. The model is based on the Boltzmann population of the bound electronic energy levels of the quasi molecule formed in the collision and the discretized continuum. The discretization of the continuum is implemented by a free electron gas in a box model assuming an effective square potential of the quasi molecule. The temperature of the electron gas is calculated by taking into account a thermodynamically adiabatic process due to the change of the effective volume of the quasi molecule as the system evolves. The system may undergo a transition with a small probability from the discretized continuum to the states of the complementary continuum. It is assumed that these states are decoupled from the thermodynamic time development. The decoupled states overwhelmingly determine the yield of the asymptotically observed fragment ions. The main motivation of this work is to describe the recently observed $H^-$ ion production in $OH^+$ +Ar collisions. The obtained differential cross sections for $H^-$ formation, cation production and electron emission are close to the experimental ones. Calculations for the atomic systems $O^+$ +Ar and $H^+$ +Ar are also in reasonable agreement with the experiments indicating that the model can be applied to a wide class of collisions.

## I. INTRODUCTION

The understanding of the basic processes induced by the collisions of ions with atoms or molecules is essential for modeling the chemical changes in various mediums, e.g., astrophysical environments [1] and living matter [2]. The heavy ions of the solar wind can be accelerated via an electromagnetic-wave–particle interaction mechanism in the magnetized plasma of the corona and reach speeds of up to several hundred km/s, which correspond to several keV energy [3]. Collisions in this energy range are important in the interaction of solar wind with cometary and planetary atmospheres and also in life sciences as they correspond to the distal region of the Bragg-peak [4]. The electron capture processes maximize here, inducing chemical changes in living tissues and astrophysical environments [5,6]. As a result of the capture and ionization processes, ionized molecules are created, which may undergo fragmentation processes due to Coulomb-explosion and form highly reactive radicals. Slow electrons are also effectively ejected from the collisions [7]. These electrons efficiently produce single- and double-strand breaks of DNA molecules due to the dissociative electron attachment (DEA) process leading to cell death in radiolysis [8]. Ordinary perturbation theories cannot handle atomic or molecular collision in this energy range. The projectile velocities are well below one atomic unit, so the perturbation is strong, which leads to difficulties in the calculations.

Recently the production of negative ions of hydrogen was observed in 7-keV $OH^+$ +Ar collisions [9,10]. The $H^-$ ions are able to form molecules in collisions with atoms due to associative electron detachment (AED). Thus they play significant role in the chemistry of planetary atmospheres and star formation as precursor of molecular hydrogen and larger molecules [11,12]. The observed H⁻ ions in Ref. [9] were due to binary collision of atomic centers detaching the H center from the OH⁺. It was unexpected that the detached center is able to grab two electrons with significant probability in those violent collisions to form the delicate loosely-bound two-electron system of H⁻. However, as it was found later, this process is general, H⁻ emission was observed from many other molecular species [9,13].

As expected, effective positive hydrogen ion formation by binary collisions was also observed [10]. The ratio of the fractions (H⁻/H⁺) in those experiments was about 1/100 and was independent of the observation angle, and thus, independent of the impact parameter of the collision. This was surprising and led to the conclusion that the different charge states were populated by some simple statistical law. This is supported by the fact that the outgoing channels for the hydrogen center are very limited, H⁺ and H⁻ have only one (stable) state, while the intermediate channels of the collisional quasi molecule are quite numerous. The large number of possible transitions between the different intermediate states may cause the process to be stochastic, which leads to the statistical distribution of the final states.

The decisive role of the electronic excitations in the process was also shown. In less violent collisions, when the impact parameter is large enough, the energy transferred to the H center is smaller than necessary to detach it from the molecular ion in its ground state. In such collisions, when the scattering angles are less than ~10°, however, large H⁻ and H⁺ yields were measured [10]. This indicates that the molecular ions are excited by the collision and a kinetic energy release (KER) separates the centers. By a semi-classical model calculation, where KER of several eV was included, the observed energy and angular distribution of the positive and negative ions could be well described [10]. Also, the recently observed large fraction of negative H⁻ fragment ions quasi-isotropically emitted from $H_2O$ target with low kinetic energies in large-impact-parameter collisions with O⁺ was reproduced by the model [13]. The main shortcoming of the model is that it does not give explanation for the formation mechanism of negative ions and the values for the relative population of the different charge states are taken from the experiment.

In Ref. [9], electrons with broad energy distribution in the 0.1-100 eV energy range were also observed. The emission shows a maximum at the low energy side and has a slowly decreasing high energy tale. This is a general feature of low energy ion-atom or ion-molecule collisions [7]. Binary collisions between the slow projectile and the target electrons cannot be the source of this emission since the energy of the recoiled electrons, which regarded as quasi free, is 0.2 eV at maximum. The observed relatively high-energy electrons may originate from multiple scattering sequentially on the target and projectile centers, however. This is the so-called Fermi-shuttle effect [14]. Up to fourfold scattering sequences for the electrons were observed experimentally in C⁺ + Xe collision in the MeV energy range [15]. The energy distribution of electrons showed wide peak structures at energies corresponding to $2V$ and $4V$ velocities, where $V$ is the velocity of the projectile. The velocity of the electrons increases by $2V$ at each scattering event by the projectile. In the keV collision energy range, these structures are smeared out, but classical trajectory calculations showed that the electrons may undergo 6–14-fold scattering sequences [16].

In this work, a model calculation of statistical type is presented in order to explain the emission of fragment ions and electrons in slow molecular collisions. The main motivation is to explain the H⁻ emission in OH⁺ +Ar collisions, for which rigorous quantum mechanical approach appears to be of prohibitive difficulty. The different electronic levels of the collisional quasi molecule OHAr⁺ are supposed to be populated by a statistical law during the collision. In this approach, transition probabilities are not involved, which greatly simplifies the calculations. The population of the excited levels of the OHAr⁺ quasi molecule that dissociate to an H⁻ ion leads to the observed anion formation. The H⁻ production cross sections resulting from the model are close to the experimental values.

The outline of this work is as follows. First, the basic concepts of the model are introduced in section II. In section III, the obtained results are presented for three collision systems in separated subsections. The results are compared with available experimental results. In section IV, the conclusions obtained by the model are summarized. In the appendices, the details of the calculations are presented.

## II. THE THEORETICAL MODEL

The model is presented for the $OH^+$ + Ar system, but it may be easily generalized to other molecular or atomic collisions. Atomic units will be used throughout the paper unless otherwise noted.

A semi-classical approach is adopted, in which the nuclei follow classical trajectories. The model uses the energy levels of the electrons, which are from *ab initio* quantum mechanical calculations or reasonable approximations. Initially, the collision partners are separated and they are in their ground state. When the projectile approaches the target, the system can be characterized by electronic eigenstates at fixed atomic centers of the quasimolecular collision system. In the course of the collision, nuclear motion couples the different states and transitions may occur between them. The outgoing channels describe excitation, capture and ionization processes. Calculating the probability of the different outcomes as a function of the impact parameter allows us to deduce cross sections for the different processes. In ordinary methods to solve the underlying problem, the expansion of the time dependent state vector of the system in the unperturbed electronic eigenstates is performed. This leads to coupled equations for the resulting time dependent coefficients of the different channels. The solution of this problem may be complicated due to the large number of channels and the strong coupling between them.

The basic concept of this model is different at this point. In the present treatment, it is assumed that the transitions between all the electronic states except some of the highly excited ones open up at a critical distance of approach $R_{crit}$, and in the further development, transitions are so frequent that the states are statistically populated. For the validity of this approximation, it is necessary that there is enough time for multiple transitions during the collision process, that is, the collision is sufficiently slow. This condition may be expressed as $R_{crit}/V \gg \tau$, where $V$ is the projectile velocity and $\tau$ is the characteristic time for transitions between the channels, which is proportional to the inverse of the square of the corresponding coupling matrix element. Since the latter one is in the order of an atomic unit as well as $R_{crit}$ for small quasimolecular collision systems, projectile velocities smaller than 1 a.u. are considered in this work.

In such slow collisions, the scattered free electrons have enough time to undergo many collisions with the atomic centers before they are emitted, giving way to frequent energy exchange between them and the atomic centers. This effect contributes to the statistical distribution of energies. The thermodynamic equilibrium is expected to be reached within a small fraction of time of the collision.

A schematic view of the collision with electric potential for the electrons is shown in Figure 1. The critical distance is taken as the classical overbarrier capture distance [17]. At this point, the most loosely bound electron may pass the barrier between the nuclei classically. Quantum mechanically, this is the typical distance, where the wavefunctions localized originally on the target or the projectile expand to both of them. This opens up transitions, since geometrical overlap of the wavefunctions increases.

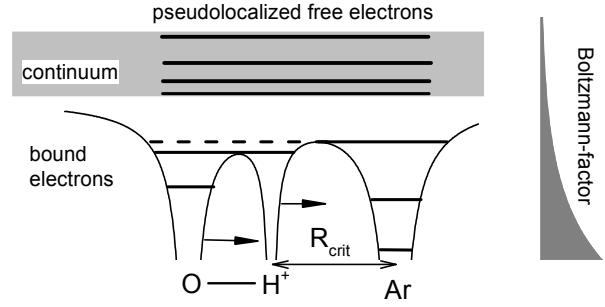

**Figure 1** Schematic view of the OH$^+$ +Ar collision system. The potential energy curve for the electrons is shown. The energy levels of electronic bound and free states are denoted by horizontal lines. At critical distance of approach the most loosely bound electron of Ar passes the barrier and becomes quasi molecular. The pseudo-localized free electrons are those, which occupy states with positive energy in the continuum, but are localized to the quasi molecule. The population of the states is depicted by the Boltzmann factor.

The critical distance according to the overbarrier model [17] is

$$R_{crit} = \frac{2\sqrt{q}+1}{-I_b}, \tag{1}$$

where $I_b$ is the binding energy of the most loosely bound electron on the neutral target, which in this case is the argon atom and $q$ is the charge state of the projectile. For the studied systems, $R_{crit}$=5.18 a. u. is obtained. The overbarrier model was developed for atomic projectile ions, where the distance of approach is measured from nucleus to nucleus. In the case of molecular projectile, this definition is ambiguous. In the model, the H center is taken to be the effective charge carrier, from where this distance is measured.

In the statistical distribution of energies, the electrons in the continuum play a significant role. In order to introduce them in this model, their energy levels have to be discretized. It is plausible to assume that only those electrons in continuum interact with the system and are involved in the statistical distribution of energies, which are localized in the volume of the quasi molecule. The energy spectrum of such electrons, which are referred to as pseudo-localized free electrons (PLFE), is discrete (see Figure 1). A derivation of their energy levels and their energy density is presented in Appendix A.

Due to the stochastic energy exchange between the nuclei and electrons, different electronic states are populated with probabilities defined by the Boltzmann factor

$$p \sim e^{-\frac{E}{kT}}, \tag{2}$$

where $E$ is the energy of the state, $T$ is the temperature and $k$ is the Boltzmann constant. Be aware that in $E$, the energy of the PLFEs as well as that of the bound electronic system is included. The resulting probabilities for populating the excited states of the quasi molecular system and the related statistical quantities like the expectation value for the energy of the system are given in Appendix B. As the collision partners approach each other, the characterizing temperature $T$ increases due to thermodynamically adiabatic compression. The equations governing the temperature are also given in Appendix B.

In the way out of the collision, the partners are getting separated and the system cools down. When the H center approaches the critical distance again, the transitions between the electronic states become less likely so that they practically stop. After asymptotic separation of the centers, certain charge states of the atomic centers may be observed depending on this final state. These are expected to be mostly low charge states since the final temperature is low. Contrary, high charge states up to threefold are observed in the experiments. Therefore one may conclude that some of the populated highly excited levels are coupled out from the stochastic time development and fixed earlier. This is also supported by the observed relatively high fraction of anions, which are accompanied by emission of cations. They are formed efficiently during the collision at close approach of the collision partners, but the system is likely to make a transition to the energetically favorable neutral–neutral or neutral–cationic state in the way out of the collision both in the picture of stochastic transitions and in coupled channel calculations. For the explanation of the observed relatively high anion production, the switching off the stochastic transitions with some probability during the collision process seems to be necessary from this point of view.

In order to explain this, the role of the active PLFEs in the stochastic transitions have to be considered. Transitions between the different excited levels of the transient quasi molecule may be hampered by the large energy difference between them. The situation is different when a PLFE is present. The energy difference can be taken by the PLFE since it can occupy many different energy levels in the quasi continuum so that the total energy of the system does not change significantly. The PLFE, however, with a small probability can leave the volume of the quasi molecule, that is, the electron is ejected. Then the electromagnetic interaction between the electron and the rest of the system becomes negligible and the energy exchange becomes practically impossible. The quasi molecule is left in a specific state, which adiabatically develops further. Its energy continuously varies with the varying internuclear distance but the jumps between different energy levels do not happen any more. In order to demonstrate this some simulations are performed, which are presented in the supplemental material [18]. The Schrödinger equation is numerically solved for a particle in a box with walls of finite heights. The width of the box slowly gets narrower in time representing the quasi molecule in the way in of the collision. If the initial state is a bound state containing only one mode with a given wavelength, that mode remains the dominant one for a time far longer that 1 atomic time unit. If another particle being in a pseudo-localized free state is also present in the box, the other modes appear very soon, indicating that transitions between the different energy levels of bound states are open.

In the present statistical model, the time development of the system is followed in small discrete time steps. In each step, the position of the nuclei, the force acting on them, the temperature of the system and the probability of populating the different excited states and PLFEs are calculated assuming statistic distribution of the excitation energies. It is assumed that when a PLFE leaves the volume of the quasi molecule the state of the remaining ionized quasi molecule is fixed as explained above, that is, a certain state is branched out from the stochastic time development. The probability of such events is calculated in Appendix C. The motion of the nuclei is then calculated by the adiabatic potential energy surface corresponding to the fixed excited state in the further time steps. Each potential surface is associated with a different dissociation limit of the constituent atomic ions at far internuclear separation. The populated charge state of the fragment ions is recorded after the collision letting the system develop for sufficient time. Many trajectories are simulated with different initial conditions. Since the time developments starting from a given initial condition is the same up to the earliest branching point, a stochastic time development is needed to be calculated only once letting it to be active for the whole time duration. Then selecting some branching points, further calculations are performed that belong to the same initial condition, in order to sample random events of branching outs.

The population of pseudo-localized free states by the Boltzmann factor and the subsequent transitions to unlocalized free states describe excitation processes to free states and ejection of electrons at each time step. The density of pseudo-localized free states obtained in Appendix A allows the determination of the energy distribution of the ejected electrons. The populated free states are expected to be s waves, since in these slow collisions, the angular momentum of the electrons is not expected to take high values. This again limits the validity of the model to low velocities. The s-waves are expected to be centered on the different scattering centers, the nuclei and free electrons. Considering this, the angular distribution of the emitted electrons can be determined (see Appendix D).

The excitation into free states may be referred to as evaporation of electrons, which appeared in the model of Russek [19] for ionization in atomic collisions. The source of the excitation energy was supposed to be a friction mechanism due to electron-electron collisions. This energy was statistically divided in uniform energy cells representing the energy levels of the atoms and the continuum. With two fitting parameters, a good agreement was obtained with the experimentally observed charge states distribution of ions following $Ar^+$+Ar collisions. In a developed version of the model, the friction mechanism was replaced by the excitation to discrete levels of autoionizing states which depends on the closest approach of nuclei [20]. Later, the model was further developed. Phase-space cells were used instead of energy cells. Using an adjusting parameter for the transitional matrix elements, a better agreement was found for the final charge state distribution [21]. Other different statistical approaches for atomic collisions were also developed [22,23].

In the present model, *ab initio* electronic energy levels of diatomic molecules are used instead of energy cells. The potential surfaces of the triatomic system are approximated by sum of diatomic potentials. Sometimes screened charged potential approximation is used if *ab initio* potentials are not available. More on the used potential curves can be found in Appendix E. In the calculations, some atomic levels corresponding to different dissociation limits are excluded. Since it is not likely that the angular momentum of the bound electrons increases in slow collisions, the atomic levels with J>4 are excluded. Energy levels lying higher than certain energy limit above the ground state, high energy cut limit (HECL) in the following, are also excluded in order to avoid population of infinite number of levels approaching the continuum. This is justified if one considers that highly excited (Rydberg) states are expanded in space, so that they cannot be considered as localized to the quasi molecule. Their overlaps with low lying states are small, therefore the transition times may exceed the collisional time. Since the characteristic radii (4 a. u.) of the 2s and 2p orbitals of hydrogen (2$l$ level in the following) are relatively large, it is assumed that the 2$l$ level should be excluded. This level is just above 10 eV excitation energy, which is, therefore, accepted as a common HECL leading to results in good agreement with the experimental ones. Unless otherwise noted, the 10 eV HECL is used in the presented results in section III. The autoionizing states are skipped merely to save computation time. Since they are highly excited, they are not effectively populated by moderate thermal processes. Despite their low population, their contributions can be observed in the experimental energy spectra, since they emit electrons in narrow peaks

No adjusting parameters like in the Russek model [19-21] are involved in the present model. The ion production and electron emission cross sections obtained for the OH$^+$ +Ar collisions, as well as for the atomic collision systems of H$^+$ + Ar and O$^+$ + Ar are in good agreements with the experiments. This indicates that the model can be applied to a wide class of collisions without system dependent adjusting parameters. Moreover, it is possible to simulate coincidence events with the model.

## III. RESULTS AND ANALYSIS

In this section the results obtained for three collision systems are presented and compared with experimental results from the literature. Since the calculations are of high computation demands, only sparse sampling in angle and energy are feasible. The theoretical results of this work are, therefore, represented by symbols, while the experiments sometimes are denoted by lines since the sampling was dense in the experiments.

### A. The OH$^+$+Ar system

The calculations for OH$^+$+Ar collisions have been performed with HECLs of 10 and 12 eV. In the former case, only the ground state is included for the atomic hydrogen, since energy of the 2$l$ level (10.2 eV) is above the limit. With the higher cut limit, the 2$l$ level is also included.

In Figure 2, the results obtained for H$^-$ and H$^+$ production cross sections using the 10 eV HECL are shown and compared with the experimental results from Ref. [9-10] for 7-keV collision energy as function of the observation angle. The calculated values agree reasonably well with the measured data. However, significant statistical errors occur in the calculation as it can be seen from the non-monotonous decrease of the obtained H⁻ production cross section above 60°. The analysis shows that the errors stems from the relatively spare sampling of the random orientation of the primary $OH^+$.

The calculation has been performed for one thousand impact parameters all with different random orientation of the primary projectile. For each impact parameter, one hundred branching outs are modeled. The process has been repeated once more. The resulting ion production cross sections for the two runs agree within a factor of 2 except for one data point. In the Figure 2, the average of the two results is shown. The statistical errors are larger than those expected from the number of the simulated events. The fact that only the initial orientation of the projectile was different in the two runs implies that the process is sensitive for the initial orientation of the $OH^+$ projectile. For atomic ion projectiles, where no orientation is involved, the statistical errors are significantly smaller as it is obvious from the obtained results shown in the next subsections.

In average, the calculations overestimate the H⁻ production cross section by about a factor of 2 and underestimate $H^+$ production by about a factor 1.5. With the 12 eV HECL, the obtained angular dependences (not shown here) are similar, but the H⁻ production is underestimated by a factor of about 8. This is probably due to the fact that the relative statistical weight for H⁻ is significantly decreased with the inclusion of the first excited state of H.

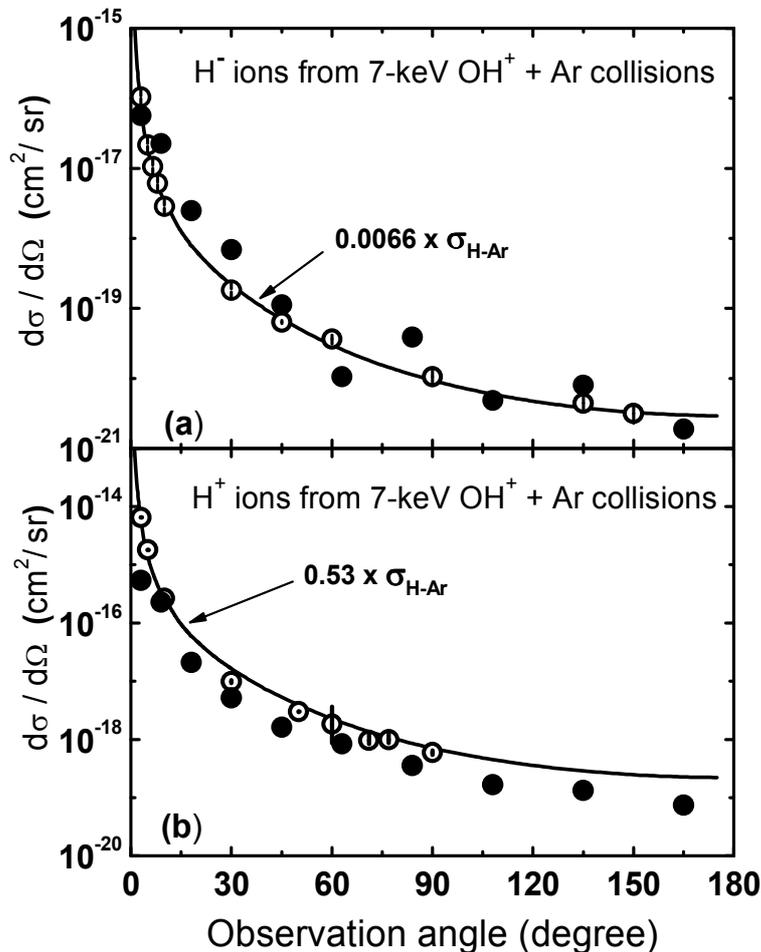

**Figure 2** Single differential cross sections according to observation angle for (a) H⁻ and (b) $H^+$ production in 7-keV $OH^+$ + Ar collisions. Closed circles correspond to this work; open circles are the experimental results from earlier works of the author et al. [9,10]; solid lines are two body H scattering calculations on Ar multiplied by fitting factors indicated in the figure, also from Ref. [9-10].

In Figure 3, the calculated electron emission cross section differential with respect to the energy and angle are shown. In the 0.2-100 eV emission energy range, the calculated results using the 10-eV HECL are in good agreement with the experimental ones [9] except in some narrow energy ranges. In the case of the 12-eV energy cut limit, the calculations resulted in values lower by a factor of up to 3.5 between 20 and 100 eV.

In order to explain the partial deviations, it should be noted that the experimental results measured by an electrostatic spectrometer have rather large uncertainties below 1 eV (not shown in Ref. [9]). There is also an instrumental effect at small observation angles below 60°, at which angles the scattered projectiles may enter the spectrometer. This leads to elevated background in the spectra, particularly visible above 100 eV, where the signal is weak.

An MNN Auger-line of Ar is prominently present between 10 and 20 eV in the experimental energy spectra. This feature cannot be reproduced with the present model since the autoionizing states are excluded from the calculation.

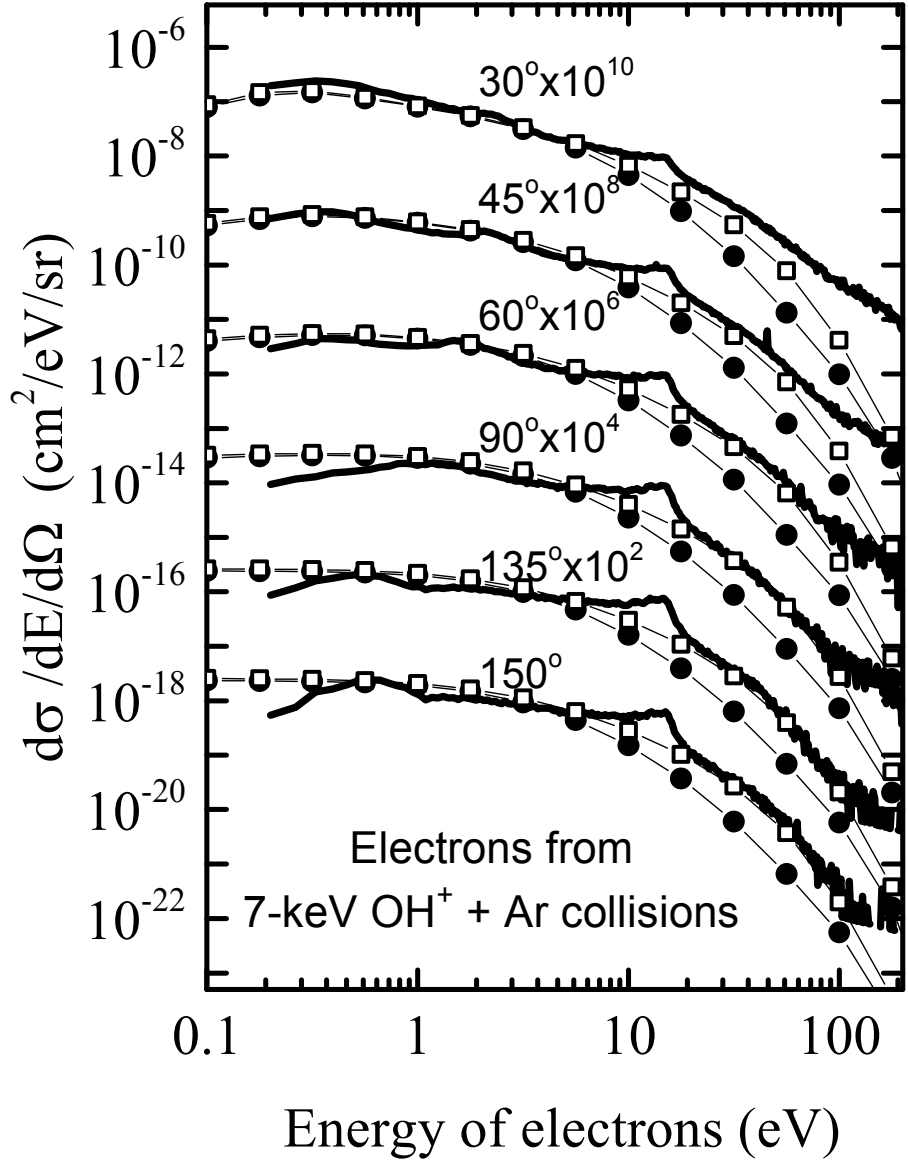

**Figure 3** Electron emission cross sections differential with respect to energy and angle for 7-keV $OH^+$ + Ar collisions at different observation angles indicated in the figure. For graphical reasons the spectra are multiplied by factors indicated next to the observation angles. Open squares: this work with HECL of 10 eV; closed circles: this work with HECL of 12 eV; solid thick lines the experimental results from the earlier work of the author et al. [9]. Symbols are connected to guide the eye.

The analysis of the trajectories shows that a large fraction of the emitted is due to such collisions where the oxygen atom suffers a close impact parameter collision with the target atom. The enhanced electron emission is due to the higher temperature of electron system reached in such collisions as it is demonstrated in Figure 4. The temperature is expressed as energy $\mathcal{E}=kT$ in units of electron volts (eV) in the following. When both projectile centers pass the target relatively far from it (a), the temperature increases up to 4 eV. In the case when the H center approaches significantly closer (b), while the oxygen is still far, the temperature moderately increases up to 6 eV. However, when the oxygen center suffers a close collision with the target (b), the maximum of the temperature reaches 28 eV. According to the calculation, in the latter cases, the ionization degree reaches 4, while in the former cases it does not exceed 1. The high temperature favors for the population of highly excited states such as the ionic $H^-–O^+$ channels. The impact parameter of the O center depends on the initial orientation of the projectile. This is the explanation why the population of ionic states is so sensitive to the orientations.

The analysis of the final charge states shows that in close collision by the O center, the most likely outgoing channel leading to $H^-$ formation is the $(e^-, H^-, O^+, Ar^{2+})$ one. In the case when the oxygen does not suffer a close collision, $H^-$ is also formed by the weaker but still significant channel, the $(H^-, O^+, Ar^+)$ one. This outgoing channel is one of those channels that couple out from the stochastic development at the critical distance and do not involve electron emission. Such channels could be identified by coincidence techniques, but no such experiments have been performed so far. A comparison of the theory with coincidence experiments may be a subject of a future work.

All together, good agreement with the experiment is obtained for this collision system in both electron and ion emission for collision velocity of 0.13 atomic unit and with HECL of 10 eV.

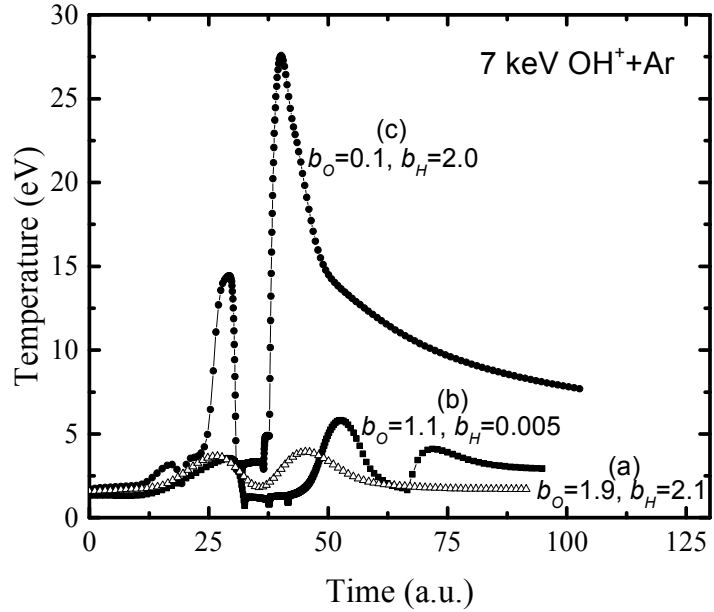

**Figure 4** Temperature of the electron system as a function of the time for 7-keV $OH^+$ + Ar collisions at different impact parameters $b_O$ for oxygen and $b_H$ for hydrogen as indicated in the figure. The starting point of the time is when the H center is at a distance of 5.18 a.u. from the target center. In the vertical axis, 1 eV corresponds to 11605 K.

## B. The $O^+$+Ar system

The calculation for this system has been performed for 50-keV collision energy. This energy corresponds to a velocity of 0.35 a.u., which is still below the upper limit of the validity of the model. In Table 1, the total experimental ion-production cross sections are compared with the calculated ones using the HECL of 10 eV. For single electron capture ($O^0$ production), they agree within the experimental uncertainties. The single electron loss ($O^{2+}$ production) cross section is underestimated by factor of 1.4. For double electron loss ($O^{3+}$ production), experimental result is not available at this energy. An estimated value is obtained from the experimental results by extrapolation in energy. The theoretical value underestimates this by a factor of about 4. This, however, does not exclude the possibility of better agreement.

Significant disagreement has been found for negative ion production, which is underestimated by more than one order of magnitude. The simplest outgoing channel leading to O$^-$ formation is the O$^-$ + Ar$^{2+}$ one. *Ab initio* potential energy curves for this channel are not available. Instead, approximate screened-charge potentials are used, which likely overestimate the exact potentials of bounding orbitals. The overestimated potential energy may explain the low population of the negative oxygen ions in the calculation.

**Table 1** Total ion production cross sections in 10$^{-16}$ cm$^2$ units. Theory this work. Experiments: O$^-$, [24]; O$^0$, [25], [26]; O$^{2+}$, [25]; O$^{3+}$, [25] (* extrapolated value).

| Charge state | Experiment | Theory |
|---|---|---|
| O$^-$ | 0.89±0.22 [24] | 0.026 |
| O$^0$ | 13±2 [25], 14.5±0.9 [26] | 14.3 |
| O$^{2+}$ | 0.16±0.03 [25] | 0.11 |
| O$^{3+}$ | 0.01 [25]* | 0.0026 |

In Figure 5 (a), the calculated differential cross sections for ion productions are presented as a function of the observation angle. It is seen that each charge state follows nearly the same angular dependence as one may expect from the statistical characteristics of the model. Differential experimental results are not available in the literature.

In panel (b) of Figure 5, the charge state distribution is shown at 30° observation angle, where also the results obtained for the HECL of 13 eV are shown. The latter results are nearly the same as those belonging to the HECL of 10 eV except for O$^0$ charge state, which is significantly enhanced. The reason behind this is that there are many levels for this charge state between 10 and 13 eV so that their inclusion increases the statistical weight for neutrals.

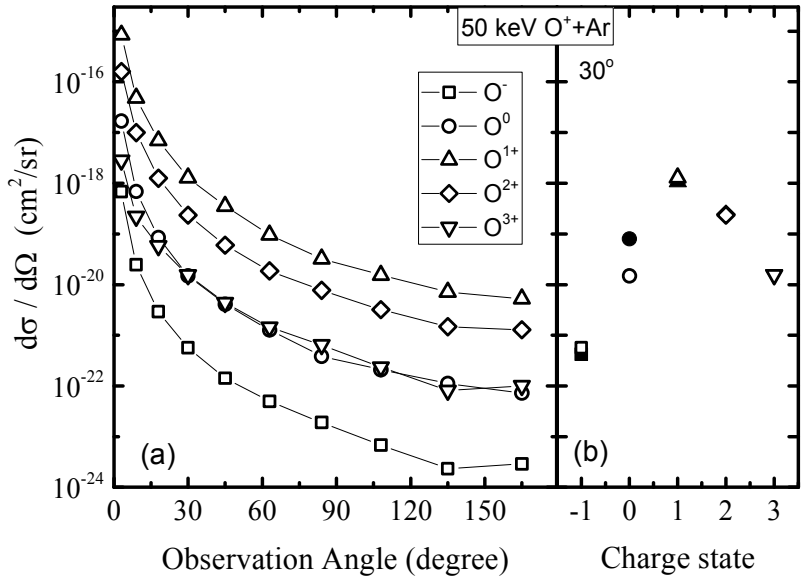

**Figure 5** Panel (a): calculated single differential cross sections according to observation angle for the different charge states of oxygen (indicated in the figure by open symbols) in 50-keV O$^+$ + Ar collisions with HECL of 10 eV. Lines are for guide the eye only. Panel (b): charge state distribution at 30°. Closed symbols are the results with HECL of 13 eV.

Double differential cross sections for electron emission are shown in Figure 6. The results of the calculations are presented for 10 and 13 eV HECLs, too. They nearly agree with each other above the emission energy of 10 eV. Below this energy, the former ones slightly exceed the latter ones. Between 20 and 100 eV, a good agreement with the experimental results [27] is found. Note that prominent Auger lines are visible between 10 and 20 eV (Ar MNN) and between 150 and 250 eV (Ar LMM) in the experimental spectra. These lines are absent in the theoretical spectra, since the model does not calculate the contribution due to the Auger process. At forward angles, the slopes of the theoretical curves are slightly smaller than the slopes of the experimental cross sections.

It should be noted that Ref. [27] reports on a remarkable feature that the cross sections slightly increase from 90° going towards larger backward angles. The origin of this feature is still under question. One key factor in this may be the fact that oxygen is significantly lighter than argon. Therefore, the oxygen center recoiled at small impact parameters moves backwards with respect to its initial velocity. In the present model, enhanced electron emission at backward angles may be expected due to the backscattered oxygen center, which grabs electrons with itself. In the results of present calculation, however, no increasing cross sections towards large angles was obtained.

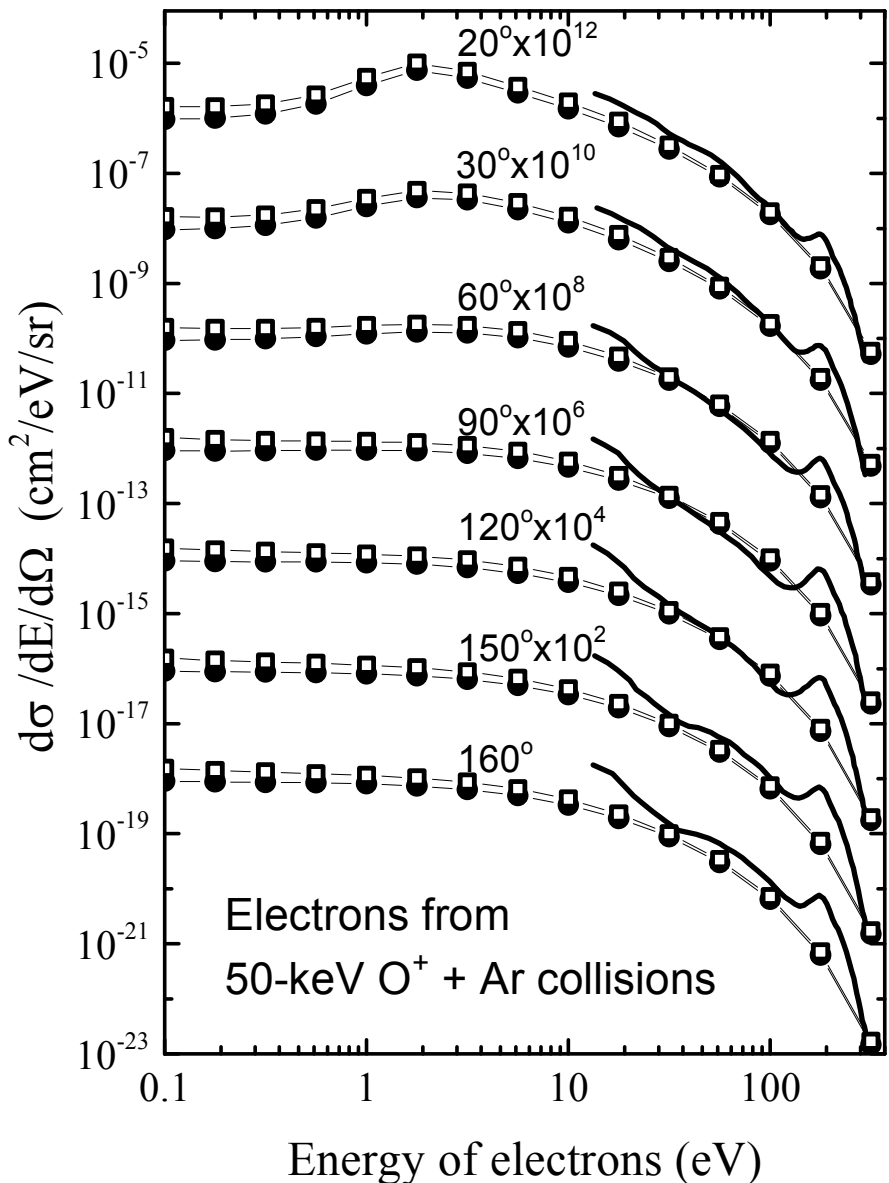

**Figure 6** Electron emission cross sections differential with respect to energy and angle for 50-keV $O^+$ +Ar collisions at different observation angles indicated in the figure. For graphical reasons the spectra are multiplied by factors indicated next to the observation angles. Open squares: this work with HECL of 10 eV; closed circles: this work with HECL of 13 eV; solid thick lines experimental results of N. Stolterfoht and D. Schneider [27]. Symbols are connected to guide the eye.

In Figure 7, a comparison with the previously investigated $OH^+$ + Ar system is made for the electron emission cross sections at 90° observation angle. Both theory and experiment show that for the atomic $O^+$ projectile, the electron emission cross sections are significantly higher than for the molecular projectile $OH^+$ at emission energies between 20 and 100 eV. One may suppose that the reason behind this is the higher collision energy. In order to check this, a calculation has been performed for $OH^+$ projectile with 53 keV, which corresponds to the same projectile velocity as that for 50-keV $O^+$. Despite the same velocity, the calculated cross sections are significantly lower in the case of molecular projectile. The results for 7- and 53-keV $OH^+$, however, nearly match. This demonstrates that the collision energy is not the decisive parameter for the intensity of electron emission. The electron emission rather depends on the type of the projectile.

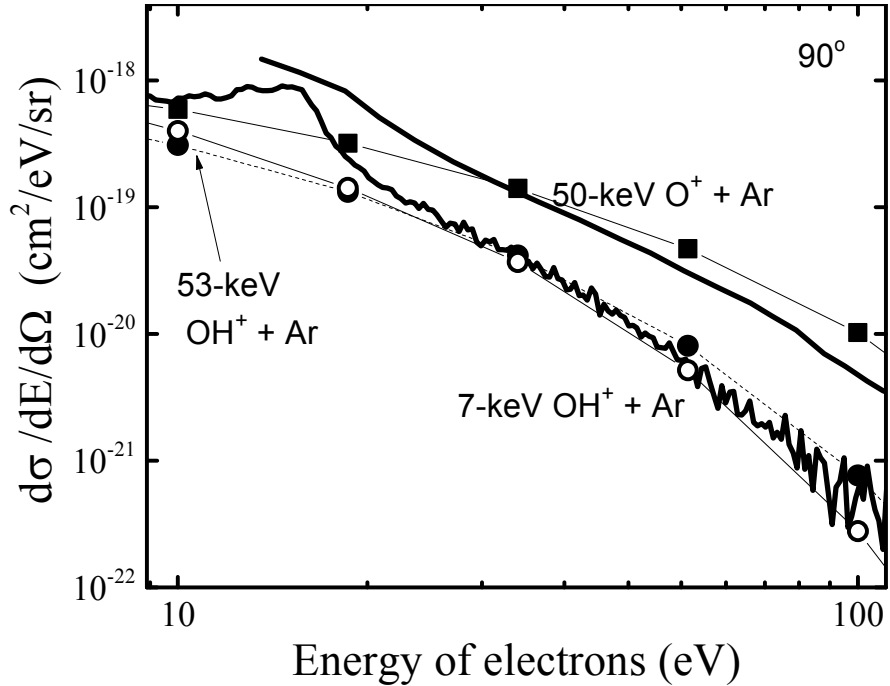

**Figure 7** Electron emission cross sections at 90° observation angle for different systems (7- and 53-keV $OH^+$+Ar and 50-keV $O^+$+Ar) indicated in the figure. Solid thick lines belong to experiments [9, 27]. Symbols denote calculations. They are connected to guide the eye. Full circles with connected with dashed lines are the results for 53-keV $OH^+$ + Ar collisions. Open circles belong to 7-keV $OH^+$ + Ar collisions, and closed squares belong to 50-keV $O^+$ + Ar collisions.

The more intensive electron emission at large emission energies in the case of atomic projectile may be caused by the higher electron temperatures. In Figure 8, it is shown that for $O^+$ + Ar collisions, indeed higher temperatures are obtained in the calculations. For a collision with impact parameter of 1 a.u. (b), the maximum of the temperature reaches 30 eV, while for the $OH^+$ + Ar system similar collisions by the oxygen leads to only 6 eV at maximum, see Figure 4(b). According to the model, the higher electron temperature in the case of atomic projectile is caused by the higher volume compression ratio of the quasimolecular system. In the model, the volume, where the electrons move quasi freely, is proportional to the internuclear distance. For atomic projectile, this can be as small as 0.1 a.u. at the closest approach. While in the case of $OH^+$ projectile, even when one of the centers approaches the target, the other center is likely to be still relatively far from the colliding centers. This practically does not allow the sum of internuclear distances to decrease below a limit, which is about twice of the bound length of $OH^+$. This limitation of the compressed volume of the quasimolecular system leads to lower adiabatic heating.

The temperature of the electron gas during the collision process also depends on the shape potential energy curves of the bound states. Therefore, the electronic structure of the colliding atoms is expected to play a significant role in the intensity of electron emission, too.

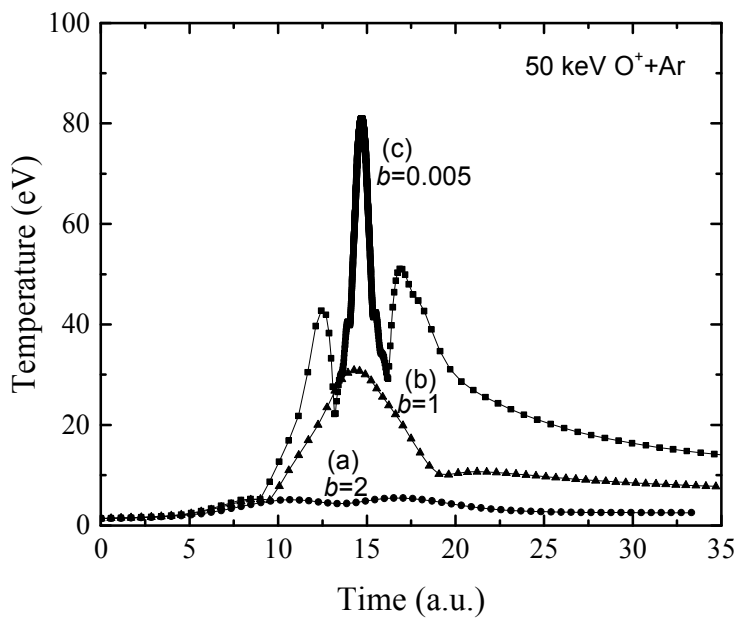

**Figure 8** Temperature of the electron system as function of the time for 50-keV $O^+$ + Ar collisions at different impact parameters *b* for oxygen indicated in the figure. The starting point of the time is when the O center is 5.18 a.u. from the target center at the beginning. In the vertical axis, 1 eV corresponds to 11605 K.

## C. The $H^+$+Ar system

Collisions for this system have been modeled with impact energies of 1 and 5 keV for which experimental data are available. These energies correspond to velocities of 0.2 and 0.45 a.u. respectively.

Single differential cross sections for $H^-$ and $H^0$ production in the 0–5° angular range are presented in Figure 9 (a) and (b) for 1 and 5 keV collision energies, respectively. The theoretical results obtained with the 10 eV HECL are shown.

For 1 keV collision energy, the theoretical result for the $H^0$ production cross section is close to experimental one [28] in the 0–1° angular range. At angles larger than 1°, the cross section is somewhat overestimated. Compared to the experimental $H^-$ cross section measured by Martinez et al. [29], the present calculation leads to an overestimation in the whole range by about a factor of 5, see Figure 9 (a).

For 5 keV collision energy (panel (b) in Figure 9), a better agreement has been found. Both the calculated $H^0$ and $H^-$ production cross section are close to the experimental ones. Significant deviations appear only above 2–3°. The bumps at 3.1° are artifact. This scattering angle corresponds to the closest approach of the nuclei of about 0.7 au. In the fitted potential energy curves there are some artificial crossings at this internuclear distance, which is close to the limit of the range of the available data. There, the potential energy curves closely approach each other so that even a small change of the potentials can lead to crossings. The calculation shows that passing through such crossings leads to increased temperature of the electronic system, causing higher populations of excited states and enhanced electron emission.

The overestimation of $H^-$ production at 1 keV may stem from the exclusion of $2l$ levels of atomic hydrogen by the HECL of 10 eV. This reduces the population of neutral states partly in the favor of negative hydrogen ion population. The inclusion of $2l$ level indeed decreases the population of $H^-$ and moderately increases that of $H^0$ as the calculations with HECL of 12 eV indicates, (not shown in the figure). The decrease of the $H^-$ cross section is, however, dramatic. It becomes about one order of magnitude smaller than the experimental one. This suggests that the 2*l* level is active, but its population is lower than that expected at thermodynamical equilibrium. It may be due to the low transition rates from the other levels. It is likely that this level, though appreciably populated, does not reach thermodynamical equilibrium during the time of the collision. For faster collisions, its population likely becomes negligible due to the shorter available time. This is supported by the fact that the theory gives nearly correct cross sections at the collision energy of 5 keV without the inclusion of 2*l* level.

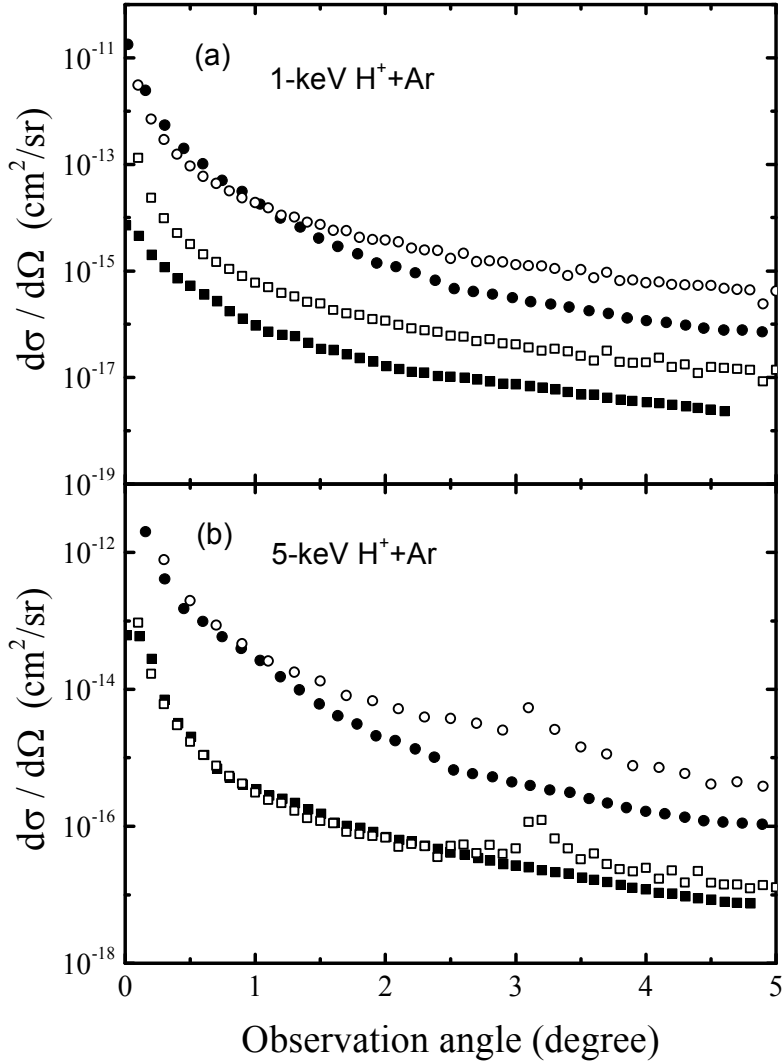

**Figure 9** Single differential cross sections according to observation angle for H⁻ (squares) and H⁰ (circles) production for (a) 1-keV $H^+$ + Ar, (b) 5-keV $H^+$ + Ar collisions. Open symbols belong to the present theory. Closed symbols denote the experimental values of Alarcon and Martinez et al. [28,29].

In Table 2, total ion production cross sections determined in Ref. [28] and [29] by integration of the measured single differential cross sections over the solid angle are compared with the results of the present calculations. The total cross sections in the present work are determined by a different method. The ion production probabilities multiplied by the cross sectional areas are integrated over the impact parameter. The so obtained cross sections are higher than the experimental ones for both H⁻ and H⁰ at 5 keV collision energy. This is somewhat puzzling, since the differential results are in good agreement, see Figure 9 b. Note that, bumps at 3.1° have little contribution to the total cross sections so that it cannot explain the difference. The high peaks at zero degree, however, have significant contribution to the total cross sections and are significantly higher in the calculations.

**Table 2** Total H⁻ and H⁺ production cross sections in $10^{-16}$ cm² units. Theory present work, experiments from Ref. [28] and [29].

| Enegy | H⁻ | | H⁰ | |
|---|---|---|---|---|
| | Theory | Exp. | Theory | Exp. |
| 1 keV | 0.15 | 0.007 | 3.8 | 7 |
| 5 keV | 0.17 | 0.04 | 20 | 13 |

In a recent theoretical work [30], total H⁰ production cross section was found to be close to the experimental values, but the obtained H⁻ production cross section significantly overestimated the experimental ones. These results were obtained by calculating the classical trajectories of the active electrons in the quasi molecule. The electrons were treated independently without interaction between them. In this classical description, the ionization potential of the H⁻ ion was significantly overestimated. This led to the overestimation of the H⁻ production. In the present model the ionization potentials of the H⁻ ion are discrete values obtained from the literature. This eliminates the drawbacks that emerge in the classical trajectory calculations. The presently obtained total H⁻ production cross sections are somewhat closer to the experimental ones. It should be noted, however, that the differential cross sections may be divergent at 0°, which makes the accurate determination of the total cross sections difficult both in theory and experiment.

Double differential cross sections for electron emission are presented in Figure 10 for three observation angles. The theoretical results obtained with the HECL of 10 eV are shown. Experimental data measured in two laboratories for this system [31, 32] are also shown in the figure. The calculated cross sections are close to the experimental ones in magnitude, but the shape of the curves are somewhat different. Significant deviations can be observed only below 1 eV at 90°. Here the experiments are less reliable, as indicated by the fact that the experimental results also deviate from each other below 5 eV. The present results do not seem to favor any of the experiments, partial agreements are found for both. Inaccuracies expected in the calculations, too, due to the artificial potential curve crossings that cause also the bumps in Figure 9 (b).

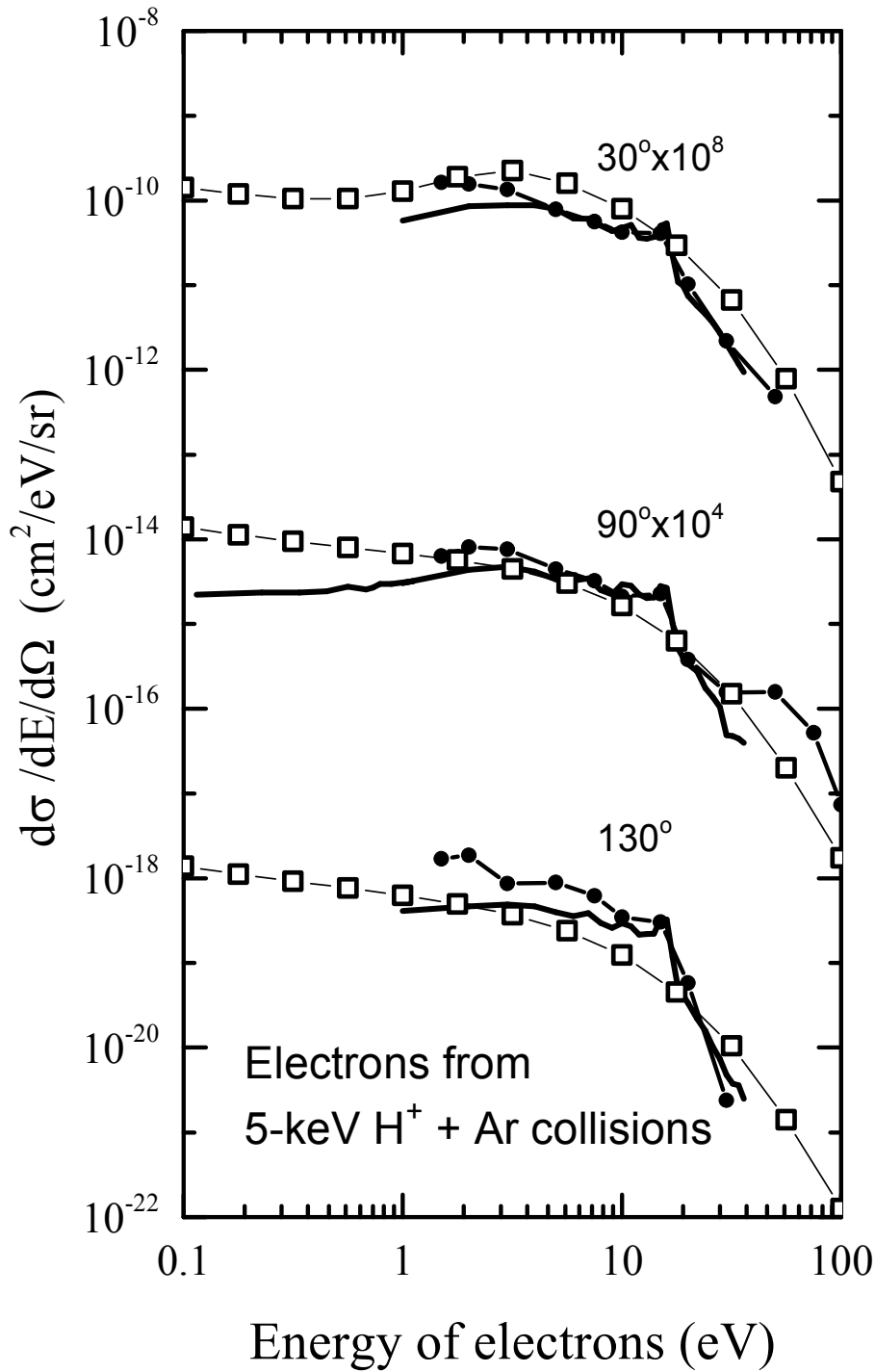

**Figure 10** Electron emission cross sections differential with respect to energy and angle for 5-keV H$^+$+Ar collisions at different observation angles indicated in the figure. For graphical reasons the spectra are multiplied by factors indicated next to the observation angles. Open squares: this work with HECL of 10 eV; closed circles: experiments at Behlen Laboratory, Nebraska; solid thick lines: experiment at Pacific Northwest Laboratory, Washington [31, 32]. Symbols are connected to guide the eye.

## IV. Conclusions

A statistical-type model is developed in order to describe the differential ion-production and electron-emission cross sections in collisions of molecular ions with atoms. The internal dynamics of the quasi molecular system formed in the collision is approximated by adiabatic heating of the electrons and the Boltzmann distribution of the electronic energy levels. The energy levels above the continuum are discretized by confining the electrons to the volume of the quasi molecule. The electrons may undergo a transition with a small probability from the discretized continuum to the states of the complementary continuum. It is assumed that due to this process the rest of the quasi molecular system is coupled out from the statistical dynamics. Population of the out-coupled states overwhelmingly determines the yield of the observed outgoing ions.

The theoretically challenging anion production from molecular cations can be handled in the model without complications. For 7-keV $OH^+$ + Ar collisions, the calculated anion production cross sections are in good agreement with the experimental data. A good performance of the model is observed also for the cation and electron emission cross sections. The obtained energy and angular dependences are very similar to the experimental ones. The angular distribution of the emitted $H^+$ and $H^-$ ions are found to be similar to each other both in the theory and in the experiment indicating that the present statistical model gives a correct account and explanation for the observed statistical behavior of final charge states found in Ref. [10]. The good agreement in ion production for small angles, which corresponds to large impact parameter collisions, where the pure binary process is not able to break the molecular projectile, indicates that the thermal process in the present model describes the disintegration of the projectile by excitations. The kinetic energy released in this process is not only due to the excitation energy, but the pressure of the quasi free electron gas contributes to it, too.

For collision systems with atomic ion projectiles of $O^+$ and $H^+$ colliding on Ar, a good agreement with the experiments is found in both dependences and magnitude. This indicates that the model can be applied for different systems. The magnitude of the cross sections is close to the experimental ones except for the anion formation from $O^+$ and from 1-keV $H^+$ projectiles. The discrepancies may partly be due to inaccuracies in the applied pair potential curves. More complete *ab initio* potential dataset is desirable to perform more accurate calculations.

It is supposed in the model that the transition rate between the different energy levels is significant only up to a certain excitation energy, and only the levels below are populated according to the Boltzmann distribution. The results with the energy limit (HECL) of 10 eV are in reasonable agreement with the experiments for almost all of investigated systems. The only exception is the 1-keV $H^+$ + Ar system, where the population of the levels just above this limit is found to be non–negligible but less than in thermodynamic equilibrium. The calculations with higher HECLs (12-13 eV), however, have led to poorer agreement for each system. A better thumb rule for the transition rates based on *ab initio* calculations and more sophisticated selection of the effectively populated states may improve the quality of the calculations.

It is shown that the temperature of the electron gas is higher for atomic ion – atom than for molecular ion – atom collisions due to the larger adiabatic compression in the case of atomic collisions. This leads to more intensive electron emission in accordance with the experiment. Also it has been found that close collisions of $O^+$ on Ar result in more intensive electron emission than the similar collisions with $H^+$ projectiles.

In the future the model may be applied for larger collision systems in interest such as $O^+$ + $H_2O$ [13]. Also, anion and cation production from carbon and metal clusters colliding on atoms attracted considerable attention recently [33-36]. The experimental results were interpreted in terms of statistical and thermodynamic models of fragment emission from the hot excited clusters after the collisions. One of the challenging tasks in those models is the determination of the excitation energy of the clusters after the collisions. The present model may be used to calculate the excitation energies of small clusters induced by atomic collisions as well as the fragmentation during the collision.

## Acknowledgments

I. Rabadan, L. Méndez and E. Bene are greatly acknowledged for calculating potential energy curves. The author is grateful to L. Gulyás L. Sarkadi, J.-Y. Chesnel and B. Sulik for carefully reading the manuscript and for valuable discussions. National Information

Infrastructure Development Institute (NIIF) is acknowledged for awarding access to resource based in Hungary at Szeged. This work was supported by the Hungarian Scientific Research Fund (OTKA Grant No. K109440).

## APPENDIX A: DENSITY OF THE STATES OF THE FREE ELECTRONS

Treating the continuum electronic states is a fundamental problem in quantum mechanical calculations. For feasible calculations, the discretization of the continuum is necessary. Several methods were developed for this purpose using the so called pseudo states. These are orthogonal states, which are constructed from finite set of basis functions [37] or obtained by the integration of eigenfunctions over disjoint intervals of the continuous positive energy spectrum [38,39]. The latter method may implement limitation on the wavefunction in space according to the Heisenberg uncertainty relation. The choice of the pseudo states is, however, not unique, which leads to uncertainties in the calculations.

In the present model, a similar but more intuitive method is followed, which gives the condition for appropriate selection of pseudo states. The method is confining the wavefunction of the free states corresponding to certain energy eigenvalues to the size of the quasi molecule. The selection of the eigenvalues is performed such that the resulting system of wavefunctions could be orthonormalized. This allows only discrete energy eigenvalues. The confining is performed by truncating the wavefunctions outside the volume of the quasi molecule. The resulting states are referred to as pseudo-localized states and the electrons being in those states as PLFEs (see Figure 1 and Figure 11). This method is justified by the following reasons: The quasi molecule acts as an effective potential valley for the electrons trying to confine them as a box. On the other hand, the electron emission is expected to proceed in two steps. Initially, the pseudo-localized states are populated, which have large overlap with the bound states.

Transitions directly to other free states, which spread in the whole space, are unlikely because of the lack of significant overlap with the bound states. The wavefunctions of the states complementary to the pseudo-localised states are truncated inside the quasi molecule. These are referred to as unlocalized free states in the following. This way the Hilbert space of free electrons is divided to an active and a non-active part, which simplifies the description of ionization processes.

The calculation of the exact energy levels of the PLFEs is complicated. Since they are expected to be spaced densely, they may be approximated by a quasi continuous distribution. The model requires only a good approximation for the density of states. It is supposed that the electrons move freely along the lines connecting the nuclear centers in an effective potential. The potential of quasi molecule is modeled by a square potential (see Figure 11). The width of effective potential well $L$ is taken to be the sum of the distance of the atomic centers plus 1 atomic unit taking into account approximately the width of the potential of the atoms $L=R_{OH}+1+R_{ArH}+1+R_{ArO}+1$. The effective depth of the potential is denoted by $V_{eff}$. As shown, one can consider the following types of wavefunctions for free electrons in such potential shape. In the general case, the wavefunctions with definite positive kinetic energies expand to the whole space and have a sinusoidal shape with different wavelength inside and outside the box, see Figure 11 (a). Due to the continuity of the derivative, the amplitude within the well may be smaller than outside. These states have small geometrical overlap with the bound states, therefore the probability of transitions between them is small. As a consequence, they are not directly populated by the thermal energy exchange. Wavefunctions of the other type are such that the part outside the box is cut down, see Figure 11 (b). These may be realized by appropriate superposition of the wavefunctions of the former type. The corresponding states, the pseudo-localized states, after normalization, have large overlaps with the bound states, so that frequent transitions are expected to occur between them. They contain many components of different wavelengths, but the expectation value of their energy is determined by the wavelength inside the box.

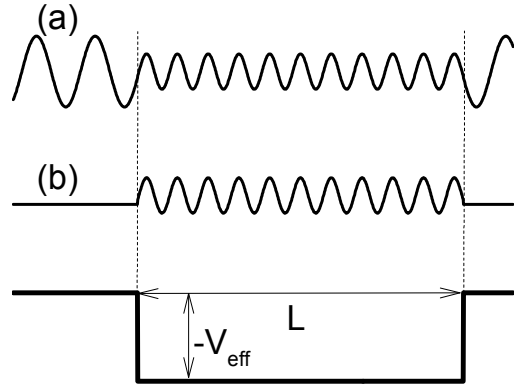

**Figure 11** The model square potential with depth of $V_{eff}$ for the quasi molecule and the type of free electronic wavefunctions. (a) General wavefunction with a definite kinetic energy. Its amplitude is smaller inside the well due to the matching the derivatives at the boundaries. (b) Pseudo-localized wavefunction.

The wavefunction solution for a square well (a particle in a box) can be written as

$$\Psi(r) = \sin\left(r\sqrt{2(\varepsilon + V_{eff})}\right), \tag{A1}$$

where $r$ is the distance from the center, $\varepsilon$ is the energy of the electron, and $V_{eff}$ is the effective potential depth. The orthogonality criterion determines the allowed wavelengths, for which

$$\frac{2\pi}{\sqrt{2(\varepsilon + V_{eff})}} = \frac{L}{N} \tag{A2}$$

is valid, where $N$ is an integer, which gives the number of energy levels up to the energy $\varepsilon$. Those energy levels, however, are not necessarily positive, so the corresponding states do not belong to the continuum. In order to have wave functions in the continuum of type (b) a common phase factor is applied to the wave functions given by (A1)

$$\widetilde{\Psi}(r) = \exp\left(-i\sqrt{V_{eff}}\,r\right)\sin\left(r\sqrt{2(\varepsilon + V_{eff})}\right). \tag{A3}$$

Due to the phase factor, the expectation values of the energies of the corresponding states are shifted,

$$\widetilde{\varepsilon} = \varepsilon + V_{eff}. \tag{A4}$$

The shifted energies start from zero value. The orthonormality of the system is conserved since the phase factor has the absolute value of unity. The density of free states is given by

$$g(\widetilde{\varepsilon}) = (N_a + n - 1)\frac{dN}{d\widetilde{\varepsilon}} = \frac{(2+n)L}{2\pi\sqrt{2\widetilde{\varepsilon}}}, \tag{A5}$$

where it is taken into account that the electrons are scattered on the $N_a=3$ atomic centers and on the other $n-1$ free electrons. With each scattering center, a series of states defined by (A2) is associated.

## APPENDIX B: STATISTICAL QUANTITIES

The method for calculating the different quantities of interest as population of energy levels, the ionization degree and the mean internal energy for the quasi molecule etc. during the thermodynamically stochastic time development is given here.

The system at ionization level $n$ is divided into two components: *1)* $n$ free electrons being in pseudo-localized states and *2)* the atomic centers with the bound electrons. The free electrons are labeled by $j$. Each of them has energy $\varepsilon_j$. The subsystem *2)* has an energy of $E_{n,i}$, where $n$ denotes the degree of ionization and $i$ corresponds to the level of excitation. The energy of the total system is

$$E = E_{n,i} + \sum_{j=1}^{n} \varepsilon_j. \tag{B1}$$

The probability of population of an excited state indexed by $n$ and $i$ is given by the Boltzmann-factor summed over all possible energy of the PLFEs. Since they follow quasi continuous distribution, integration is used instead of a sum:

$$p_{n,i} \sim \frac{1}{n!} \int \cdots \int g(\varepsilon_1) \cdots g(\varepsilon_n) e^{-\frac{E_{n,i} + \sum_{j=1}^{n} \varepsilon_j}{kT}} d\varepsilon_1 \cdots d\varepsilon_n, \tag{B2}$$

where $g(\varepsilon_j)$ is the density of states given by (A5). The division by $n!$ is applied because all possible permutations of $\varepsilon_j$'s establish the same state since electrons are indistinguishable. It is necessary for the validity of (B2) that the energies of the free electrons should be independent. This is a plausible condition, since except binary collisions free electrons weakly interact with each other. The effects of Pauli's exclusion principle are neglected here, since the occupancy of the pseudo states is expected to be much smaller than unity. Simplification and normalization of expression (B2) leads to

$$p_{n,i} = \frac{1}{Z_{tot} n!} e^{-\frac{E_{n,i}}{kT}} Z^n, \tag{B3}$$

with the state sum for the electrons

$$Z = \int_0^\infty g(\varepsilon) e^{-\frac{\varepsilon}{kT}} d\varepsilon, \tag{B4}$$

and with the total state sum

$$Z_{tot} = \sum_n \sum_i \frac{1}{n!} e^{-\frac{E_{n,i}}{kT}} Z^n, \tag{B5}$$

which acts as a normalizing factor.

It should be noted, if instead of the pseudo-localized states, all free states were taken into account, then $g(\varepsilon)$ and thereby Z, too, would be divergent. This shows that for a realistic theory, discretization of the free states is necessary.

Using (A5) an analytic expression for Z

$$Z = \frac{L(2+n)}{2\sqrt{2\pi}} \sqrt{kT} \tag{B6}$$

is obtained from (B4).

The probability that $n$ free electron is populated, or in other words, the probability of n-fold ionization is

$$p_n = \sum_i p_{n,i} \tag{B7}$$

These probabilities sum up to unity

$$\sum_n p_n = 1, \tag{B8}$$

as follows from Eqs. (B3) and (B5).

The expectation value of the ionization degree is

$$\bar{n} = \sum_n n p_n. \tag{B9}$$

The mean energy of a free electron is

$$\bar{\varepsilon} = \frac{\int_0^\infty \varepsilon g(\varepsilon) e^{\frac{\varepsilon}{kT}} d\varepsilon}{Z} = \frac{kT}{2}. \tag{B10}$$

This is the usual expression in equipartition theorem for the average kinetic energy per degree of freedom. It should be noted that in the model the motion of the electrons is restricted along the lines connecting the nuclei. Therefore the electrons have one degree of freedom.

The internal energy of the system is the mean value of the energy,

$$U = \sum_n \sum_i p_{n,i}(E_{n,i} + n\bar{\varepsilon}). \tag{B11}$$

$E_{n,i}$ depends on the internuclear separation as well as $p_{n,i}$, which also depends on the temperature.

In order to determine the change of the temperature as function of time the motion of the nuclei is followed in small time steps defined later. First, the change of the temperature of the free electrons (subsystem 1) is determined when the internuclear separation coordinates $R_1$ is changed to $R_2$ by a small step while switching off the interaction with the rest of the system. $R$ is a general coordinate including all the internuclear separation distances $R=(R_{OH}, R_{ArH}, R_{ArO})$. According to thermodynamics, the free electron gas adiabatically heats up as the volume $V_1$, where the electron can freely move is decreased to $V_2$. The initial temperature $T_1$ increases to

$$T' = T_1 \left(\frac{V_1}{V_2}\right)^{\frac{2}{f}}, \tag{B12}$$

where $f$ is the degree of freedom of the electrons, which equals to 1, since the motion of the electrons is essentially one dimensional in the quasi molecule. The volumes $V_{1,2}$ are supposed to be proportional to $L_{1,2}$, which corresponds to the picture that the electrons can move freely along the lines connecting the nuclear centers. The volume ratio therefore can be expressed with the internuclear distances. The mean energy of the electrons increases due to the increase of the temperature. The energy of the rest of the system also changes due to the changes in the potential energies. The populations of the energy levels of subsystem 2) are kept fixed as a first order approximation. The total internal energy after the small change of the coordinates of the nuclei therefore becomes

$$U_2 = \sum_n \sum_i p_{n,i}(R_1, T_1)\left[E_{n,i}(R_2) + n\bar{\varepsilon}(T')\right] \tag{B13}$$

This formula does not correspond to a thermal distribution in equilibrium (see the different temperatures in it). Switching the energy exchange on while the nuclei are kept fixed, this internal energy is redistributed according to a new temperature $T_2$

$$\sum_n \sum_i p_{n,i}(R_2, T_2)\left[E_{n,i}(R_2) + n\bar{\varepsilon}(T_2)\right] = U_2. \tag{B14}$$

This equation determines the new temperature. Since the equation is transcendent, $T_2$ has to be determined numerically at each step of the simulation.

The force $\underline{F}_X$ that acts on the atomic center $X$ and governs its motion while the stochastic energy exchange is active can be determined from the internal energy $U$. The latter quantity can be regarded as the average potential energy of the system therefore

$$\underline{F}_X = -\frac{dU}{d\underline{R}_X}. \tag{B15}$$

, where $\underline{R}_x$ is the location vector of the atomic center $X$. Tree-component vectors are underlined throughout the text. Note that, $U$ contains the energy resulting from nucleus–nucleus interactions, too, since $E_{n,i}$ is the total energy of the bound subsystem. After some math using the expression (B13) for $U_2$, which gives the internal energy $U$ as function of varying nuclear coordinates $R_2$

$$\underline{F}_X = -\sum_n \sum_i p_{n,i} \left[\frac{dE_{n,i}}{dR_2} + n\frac{d\bar{\varepsilon}(T')}{dR_2}\right]_{\substack{R_1=R,\\R_2=R}} \frac{dR}{d\underline{R}_X} \tag{B16}$$

is obtained. The first term in the bracket gives the average of the forces corresponding to the potential energy surfaces of the different states. The second term can be identified as a force due to the pressure of the free electron gas. Since the population of the energy levels is statistical, some statistical deviations from the resulting trajectories of the atomic centers are expected. However, it is expected that the rapid change of the populations averages out the statistical fluctuations.

The trajectories of the atomic centers while the stochastic energy exchange with the electrons is active are calculated by a finite step method of the Newtonian equations

$$M \frac{d^2 \underline{R}_X}{dt^2} = \underline{F}_X. \qquad (B17)$$

Adaptive time steps by the practical formula

$$\Delta t = \frac{1}{5\sqrt{\sum_X \left|\dot{R}_X\right|^2} + 2500\sqrt{\sum_X \left|\ddot{R}_X\right|^2}} \qquad (B18)$$

are used. The time steps are finer where the velocity or the acceleration of the centers is larger. They vary between 0.01 and 2 atomic units. Further refining of the time steps did not change the trajectories significantly indicating that the accuracy of the calculation with the present time step determination is sufficient. At each step, the coordinates and their derivatives, the temperature, the ionization degree and the population of charge states are calculated and recorded. Initially the charge carrier H center is placed at $R_{crit}$ distance from the Ar center with different impact parameters $b$ ranging from zero to $R_{crit}$. The position of the O center is at 1.95 a.u. distance (which is the relaxed bond length) from the H center with random orientation. Typical trajectories are show in Figure 12.

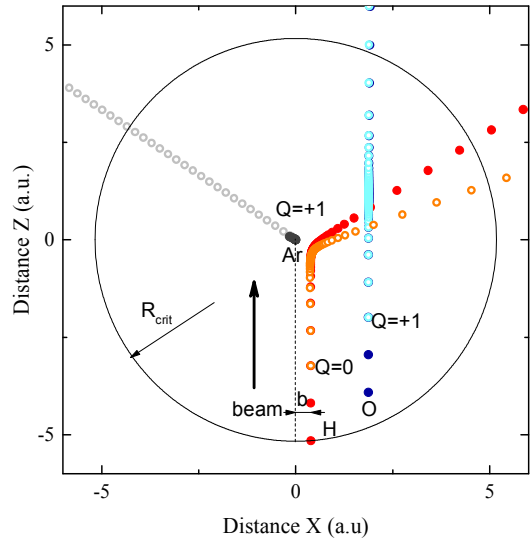

**Figure 12** (Color online) Trajectories of the atomic centers. The positions of the centers are shown by dots at different time steps of the calculation: for H (red and orange); for O (blue and cyan); and for Ar (dark grey and grey). The closed symbols (first colors) correspond to the stochastic development, while open symbols (the second colors) correspond to the development after branching out with fixed charge states Q=(0, +1,+1) for (H, O, Ar) centers respectively.

The initial temperature of the system is determined by the initial energy of the system if it is in an excited state. It should be noted that the ground state corresponds to zero temperature, which is not allowed if one concerns a thermodynamically adiabatic process. Therefore, if the system is in the ground state by the ansatz, the internal energy is taken to be that of the lowest excited state assuming that this is the most likely state to be populated when the transitions open up at nuclear separation $R_{crit}$. This initial energy is then thermalized so that the expectation value of the internal energy should be equal to it. The initial temperature is numerically determined by the equation

$$E_{init} = U(T,R). \qquad (B19)$$

In Figure 12, example trajectories corresponding to the stochastic development of the system and also corresponding to the time development after branching out are presented. At branching out, one of the excitation levels of the quasi molecule is fixed with some probability. It is impossible to perform calculations for all the possible branching out, but with sufficient number of sampling, a good account for the development of the most likely outgoing channels can be achieved. With a calculated probability given by the next appendix, a time step is selected when the branching out occurs and also a fixed excitation level is selected by the probability (B3). The subsequent motion of the centers is calculated according to the force given by the corresponding potential energy surface of the selected excitation level

$$\underline{F}_{X,n,i} = -\frac{dE_{n,i}}{dR}\frac{dR}{d\underline{R}_X}. \qquad (B20)$$

This process of simulated branching outs is repeated several times in order to collect enough data for good statistics of the production of the most likely outgoing atomic centers. The O and H centers are taken as separate atoms or ions if the distance between them is larger than 9 a.u. at the end of the calculation (200 atomic time unit after the branching out). The single differential cross sections (SDCS) with respect to observation angle $\theta$ of the regarded ionized atomic centers at given charge state is calculated as

$$\frac{d\sigma}{d\Omega}(\theta) = \sum_j \frac{b_j \Delta b N_j(\theta, \Delta\theta)}{M_j \sin(\theta) \Delta\theta}, \qquad (B21)$$

where $M_j$ is the number of simulated branching outs for an impact parameter $b_j$, and $N_j(\theta, \Delta\theta)$ is the number of events in which the regarded ion is emitted with an angle falling into the binning interval $\pm\Delta\theta/2$ around the observation angle $\theta$. Sometimes a non-uniform binning is used in order to have sufficient statistics at each angle.

## APPENDIX C: PROBABILITY OF BRANCHING OUT

As model conjeture, the fixing of a certain state of the quasi molecule happens when at least one electron escape from it. This process may be identified as a transition to an unlocalized free state. It is not feasible to calculate by strict quantum mechanics the probability of such transitions. But good approximation can be obtained classically. It is assumed that initially, the free electrons are evenly distributed within the size of the quasi molecule, which is modeled now by a sphere for simplicity with radius of $r=(L/3)$ (see Appendix A). The thermally distributed free electrons have a velocity distribution of

$$f(\underline{v}) = \frac{g(\varepsilon(\underline{v}))e^{-\frac{\varepsilon(\underline{v})}{kT}}}{4\pi v Z}. \qquad (C1)$$

Here spherical symmetry is assumed, since only s-waves are expected to be populated. Furthermore, the velocities of the atomic centers, which may shift the centers of these distributions, are neglected here, since they are much less than 1 a.u. compared to the velocities of free electrons typically of several atomic units. The number of electrons which escape from the confining sphere during a small time interval $\Delta t$ is calculated by elementary considerations.

The number of electrons leaving the quasi molecule in one direction with velocity v±$d$v/2 is

$$dN_e = dv \frac{3\bar{n}\Delta t}{4\pi r^3} \int_0^{2\pi}\int_0^{\pi/2} vf(v)\cos(\theta) r^2 \sin(\theta) d\theta d\varphi = vf(v) dv \frac{3\bar{n}\Delta t}{4r}. \qquad (C2)$$

The expectation value of the total number of escaping electrons may be expressed by the following integral

$$N_e = \int_0^{2\pi}\int_0^{\pi/2}\int_0^{\infty} \frac{dN_e}{dv} v^2 \sin(\theta) dv d\theta d\varphi, \qquad (C3)$$

which is integration of the differential quantity $dN_e$ over all velocities in every directions. The result is

$$N_e = \frac{3\bar{n}\sqrt{kT}\Delta t}{2\sqrt{2\pi}r}. \tag{C4}$$

The probability of branching out at time step *l* under the conditions that no branching occurred before is equal to the expectation value of the number of escaping electrons. Note that $N_e$ is also a differential quantity since it corresponds to a small time interval $\Delta t$, so that it is expected to be much smaller than 1. The probability $P_l'$ that branching out does not occur before time step *l* is

$$P_l' = \prod_{j=1}^{l-1}(1-N_e(j)). \tag{C5}$$

And the probability of branching out $P_l$ to occur at time step *l* is

$$P_l = N_e(l)P_l'. \tag{C6}$$

The final step *m* has to be treated separately. At the final step, the stochastic time development stops any way, because the charge carrier of the projectile reaches critical distance so that all transitions are closed. Therefore, all the states are fixed at the final step even if no electron escapes from the quasi molecule. The probability for this to happen is $P_m = P_m'$.

After a time step is selected, when branching out occurs, the question still remains which state is fixed. In order to calculate the corresponding probabilities, besides population probabilities (B3), it has to be taken into account how many PLFEs are associated with the states. This is given by the ionization degree *n*. The probability of branching out is linearly scales with it, therefore, for a given state, the fixing probability is

$$p'_{n,i} = \frac{np_{n,i}}{\sum_{n,i} np_{n,i}}. \tag{C7}$$

At final time step, the formula (B3) is applied for the fixing probability of given state since it is no more connected with electron emission.

## APPENDIX D: ANGULAR DISTRIBUTION OF THE EMITTED ELECTRONS

According to the experiments, the angular distribution of the electrons is enhanced in the forward direction. In the framework of the model, this is due to the motion of the atomic centers on which the velocity distribution of the scattered electrons is centered. It is assumed that on each scattering center there is a free electron population with equal probability. The scatterings result in zero angular momentum electrons (s-waves) in most of the cases, since the velocity of the scattering centers is well below an atomic unit and the effective impact parameter of the electrons is limited to a few atomic units. This gives a net angular momentum for electrons close to zero with respect to the scattering center. The wavefunctions of these states far from the scattering center inside the box behave as that of a free particle. In case of electron-electron collision, considering the equal masses, the position of the scattering center is taken to be the center of their mass. The energy distribution of the free electrons is determined by the Boltzmann-factor. At a given moment of the collision when *n* PLFEs are present, their velocity distribution in the laboratory frame is given as follows

$$F_n(\underline{v}) = \frac{n}{n-1+N_a}\left(\sum_X f(\underline{v}-\underline{v}_X) + (n-1)\int \frac{F_n(\underline{u}_1)}{n}\frac{F_n(\underline{u}_2)}{n} f\left(\underline{v}-\frac{\underline{u}_1+\underline{u}_2}{2}\right)d\underline{u}_1 d\underline{u}_2\right). \tag{D1}$$

This is a self-consistency equation for the velocity distribution of *n* electrons $F_n(\underline{v})$. The velocity distribution of the single electrons *f* is defined in Eq (C1), $\underline{v}_X$ and $\underline{u}_{1,2}$ are the velocities of the scattering atomic centers and electrons, respectively. The sum that appears first in the parenthesis in (D1) describes the electrons scattered on the atomic centers. The single electron velocity distributions are shifted by the velocities $\underline{v}_X$ of the atomic centers *X* and summed. The following term describes electron-electron scattering in which the energy exchange with the rest of the system is involved. Before the scattering, the electrons have $\underline{u}_1$ and u₂ velocities. After the scattering event the one of the electrons will have velocity of v with a distribution of $f$ centered on the center of mass of the electrons. This distribution is integrated over all the possible velocity combination of colliding electrons weighted by their velocity distribution and multiplied by the number of the scattering electrons $n$-1. In order to get normalized $n$-electron distribution, the whole expression is multiplied by the total number of the free electrons $n$, and divided by the total number of the scattering centers. The number of the scattering centers is the number of the atomic centers $N_a$ plus the number of the scattering electrons $n$-1.

In order to solve Eq. (D1) for $F_n(\underline{v})$, a Fourier transformation is used that transforms (D1) into algebraic expression

$$\widetilde{F}_n(\omega) = \frac{n}{n-1+N_a}\left(\widetilde{f}(\omega)\sum_X e^{i2\pi\omega\cdot\underline{v}_X} + \frac{(n-1)}{n^2}\widetilde{f}(\omega)\left(\widetilde{F}_n(\omega/2)\right)^2\right). \quad (D2)$$

Analytic solution cannot be found for this equation, since the argument of $\widetilde{F}_n$ is divided by 2 on the right hand side. However, good approximation can be obtained by an iterative method. In first order approximation, the second term is neglected

$$\widetilde{F}_{n,1}(\omega) = \frac{n}{n-1+N_a}\left(\widetilde{f}(\omega)\sum_X e^{i2\pi\omega\cdot\underline{v}_X}\right). \quad (D3)$$

In higher orders, the lower order approximation is substituted in the right hand side, which gradually gives a better approximation

$$\widetilde{F}_{n,i}(\omega) = \frac{n}{n-1+N_a}\left(\widetilde{f}(\omega)\sum_X e^{i2\pi\omega\cdot\underline{v}_X} + \frac{(n-1)}{n^2}\widetilde{f}(\omega)\left(\widetilde{F}_{n,i-1}(\omega/2)\right)^2\right). \quad (D4)$$

In the calculation, the 5$^{th}$ order approximation is used. The obtained algebraic expression is too complicated to perform inverse-Fourier transformation on it analytically. Therefore approximate solution for $F_n(\underline{v})$ is obtained by numeric inverse-Fourier transformation.

$$F_n(\underline{v}) \cong \mathcal{F}^{-1}\left(\widetilde{F}_{n,5}(\omega)\right) \quad (D5)$$

In order to get the overall velocity distribution of the electrons at a given moment of the collision, the results obtained in this way are weighted by the ionization probability and summed over all ionization degree

$$F(\underline{v}) = \sum_n p_n F_n(\underline{v}). \quad (D6)$$

The distribution of the observable electrons, which are escaped from the quasi molecule, is calculated by taking into account the escape probabilities. Electrons of higher velocities can escape with higher probability from the quasi molecule. The velocity distribution of the observable electrons is obtained similar to Eq. (C2)

$$\frac{dN_e}{dv} = \frac{3\Delta t}{4\pi r^3}\int_0^{2\pi}\int_0^{\pi/2} vF(\underline{v})\cos(\theta)r^2\sin(\theta)d\theta d\varphi = vF(\underline{v})\frac{3\Delta t}{4r}, \quad (D7)$$

except that here instead of $f(v)$ the velocity distribution that takes into account the effect of the moving scattering centers $F(\underline{v})$ is used. The cumulative velocity distribution of the observed electrons associated with a trajectory is given by the sum of partial contributions of each time step taking into account the probability of the branching outs as follows

$$F_{traj}(\underline{v}) = \sum_l^{m-1}\left(P_l'\frac{dN_e}{dv}(l) + P_l\sum_n \frac{np_n}{\bar{n}}\frac{n-1}{n}F_n(\underline{v},l)\right) + P_m'F(\underline{v},m). \quad (D8)$$

The second term in the sum represents a contribution from multiple ionization processes where more PLFEs are involved above the electron, which escapes within a selected time step causing the branching out. The corresponding states of excess PLFEs are expected to be fixed by the branching out as well as the bound molecular states. The PLFEs are finally released as the collision partners separate. This term represents a small contribution, since multiple ionization processes are not dominant. But with its inclusion, a better agreement with the experiments has been found.

Absolute cross sections for the distribution of emitted electrons are obtained by performing the calculation for different impact parameters. The cylindrical symmetry of the system is used for sparing computation time. The distribution of electrons associated with one impact parameter is rotated around the Z-axis with finite steps of 10° and averaged. This may cause deviations from the calculations performed ideally with random initial position and orientation for the projectile. Since the used set of the impact parameters and rotations is dense enough, those effects are expected to be small. The double differential cross sections (DDCS) with respect to energy and angle are calculated as

$$\frac{d\sigma}{d\Omega dE} = \frac{v \sum_j 2\pi b_j \overline{F_{traj}(\underline{v})}_j \Delta b}{4\pi}, \quad (D9)$$

where $\overline{F_{traj}(\underline{v})}_j$ are the rotationally averaged velocity distributions for the different trajectories.

## APPENDIX E: ENERGY LEVELS OF THE QUASI MOLECULE

The energy levels of the triatomic system $E_{n,i}$ at ionization level $n$ depend on all the three internuclear distances. Strict quantum mechanical calculations for the 3-dimensional potential energy surfaces would lead to high computational demand. Instead the sum of pairwise potentials of the binary subsystems (OH, ArO, ArH) is used as an approximation. At asymptotic internuclear separation the energy of the quasi molecule is the sum of atomic energy levels, which are taken from the NIST database [40].

The labeling of $E_{n,i}$ is abbreviated. To fully characterize the energy levels the index $n$ is extended to set of indices ($nh, no, na$) denoting the ionization degree of the H, O and Ar atoms respectively. Whenever sum over $n$ is performed in the statistical quantities in Appendix B, it denotes the sum over all three indices taking into account the $n=nh+no+na-q$ relation. Likewise, the level index $i$ should be extended to ($ih, io, ia$) denoting the excitation level of the H, O and Ar respectively. The energy level of the system is expressed as sum of atomic energy levels and pairwise potentials

$$E_{nh,no,na,ih,io,ia} = E^H_{nh,ih} + E^O_{no,io} + E^{Ar}_{na,ia} + P^{OH}_{no,io,nh,ih}(R_{OH}) + P^{ArO}_{na,ia,no,io}(R_{ArO}) + P^{ArH}_{na,ia,nh,ih}(R_{ArH}), \quad (E1)$$

where $E^H_{nh,ih}$ is the energy of atomic H at ionization degree $nh$ and at excitation level $ih$, and $P^{OH}_{no,io,nh,ih}(R_{OH})$ is the potential energy of OH as a function of the internuclear distance $R_{OH}$ with dissociation limit of O at ionization degree $no$ and at excitation level $io$ and H at ionization degree $nh$ and at excitation level $ih$. The meaning of other terms is straightforward from the indices. The pairwise potentials are taken from the literature if available and are leveled such that they asymptotically approach zero. Where data are not available screened charge Coulombic potentials of the form

$$P^{OH}_{no,io,nh,ih}(R_{OH}) = \frac{q^H_{nh}(R_{OH}) q^O_{no}(R_{OH})}{R_{OH}} \quad (E2)$$

are used as an approximation. In this approximation the shape of the potential curve does not depend on the excitation levels of the atoms or ions. The screening of the charge distributions are taken to be exponential

$$q_{nh}^{H}(R) = \begin{cases} \text{if } nh = -1, -1 + e^{-R/3.9} + e^{-R} \\ \text{otherwise}, nh + (1-nh)e^{-R} \end{cases}$$

$$q_{no}^{O}(R) = \begin{cases} \text{if } no = -1, -1 + e^{-R/2.6} + 8e^{-R} \\ \text{otherwise}, no + (8-no)e^{-R(no+1)} \end{cases} \quad (E3)$$

$$q_{na}^{Ar}(R) = na + (18-na)e^{-R(na+1)}.$$

The effective radii of the screenings are set to be inversely proportional to the ionization degree plus 1. In the case of negative charge states the negative-ionic radii from the literature are taken as the effective radii. In the statistical quantities the degeneracy of the $E_{nh,no,na,ih,io,ia}$ levels are taken into account by weighting the corresponding Boltzmann factors with

$$G = (2J_{nh,ih}^{H} + 1)(2J_{no,io}^{O} + 1)(2J_{na,ia}^{Ar} + 1), \quad (E4)$$

where $J$ denotes the total angular momentum of the atomic ions at the dissociation limit. Above ionization degree of 1 of both partners, both screening radii are small and the ions can be considered as point like charges. Therefore, at high ionization degrees, the energies obtained from screened Coulombic potentials are expected to be close to the real ones. For low ionization degrees, where molecular features are pronounced in the potential energy curves, the use of *ab initio* potentials is essential for accurate results.

The potential energy curves taken from the literature are fitted by approximating functions. Table 3 gives a list of the data sources and fitting functions with descriptions of their usage. In a general case, different potential energy curves exist for a given dissociation limit of a diatomic system. These are labeled by the molecular terms. A molecular term can be degenerated. In this case, replicas of that term are introduced in the calculation to take into account its multiplicity.

The Boltzmann factors associated with a given atomic dissociation limit are averaged over all the combination of the diatomic molecular terms and weighted by the factor (E4) in the calculated probabilities. Arbitrary combinations of the three diatomic molecular terms may overestimate the number of possible states, but (E4) gives the correct weights. In the case when the model is applied to collisions between atomic ions and atoms, such overestimation does not occur.

The fitting functions for the potential curves have the form of

$$f_{nx,ix,ny,iy}^{XY}(R) = P_{nx,ix,ny,iy}^{XY}(R) + A\left((1-e^{-aR+\ln 2})^2 - 1\right) + B\left((1-e^{-bR+\ln 2})^2 - 1\right), \quad (E5)$$

which consist of a screened potential given by Eq. (E3) in order to ensure the correct asymptotic behavior, and of several Morse potentials, in most cases two, in order to achieve good fit at molecular distances with $A$, $a$, $B$ and $b$ as fitting parameters. The available data do not always cover the internuclear distance range that occurs in the calculations. Since the fitting functions contain the screened potentials, more or less correct behavior is expected in the non-covered range, too.

In Figure 13, some of the resulting fit functions for the overall three center system (Ar$^+$(ground state)+OH) are shown compared with the data from the literature as function of the OH distance. When the Ar$^+$ center is asymptotically far in panel (a), the fit functions reproduce the two body OH potential energy curves from the literature. In this case, the separate Ar$^+$ ion only implies a shift in the total energy of the system. In panel (b), the case is presented when the Ar$^+$ center is at a molecular distance of 2.5 au. from the H center. The potential curves are significantly changed. Some of the ionic states (red curves in electronic version) get closer to the levels corresponding to atomic dissociation limits (black curves in electronic version), and even cross them at large internuclear distances of 3 atomic units. As the energy of the levels of the ionic states are getting lower relatively to the other levels, the population of the ionic states increases according to the Boltzmann factor. This effect enhances the positive and negative ion emission in a triatomic collision system. It is to be also noted that the diatomic potential energy curves splits due to the vicinity of the third center in Figure 13 (b).

**Table 3** List of diatomic potential data and their usage in the calculations with fitting functions [41-56].

| binary subsystem | dissociation limit | mol. term | data source | R range (a.u.) | lable of fit function | used potentials for each limit | Comment |
|---|---|---|---|---|---|---|---|
| OH | O($2s^22p^4$; $^3P_2$)+H(1s; $^2S$) | $X^2\Pi$ | [41] | .13 -10 | fOH1 | {2xfOH1, 2xfOH2, fOH3} | |
| | | $a^4\Sigma^-$ | | 1.5 -6.7 | fOH2 | | |
| | | $1^2\Sigma^-$ | | 1.55 -7 | fOH3 | | |
| | O($2s^22p^4$; $^3P_{0,1}$)+H(1s; $^2S$) | $1^4\Pi$ | [42],[43] | 1.73 -7 | fOH4 | {fOH4} | |
| | O($2s^22p^4$; $^1D$)+H(1s; $^2S$) | $A^2\Sigma^+$ | | 1.21 -7 | fOH5 | {fOH5, 4xfOH5a} | |
| | | $2^2\Pi$ | [44] | 1.56-9.57 | fOH5a | | |
| | | $^2\Delta$ | | — | | | |
| | O($2s^22p^4$; $^1S$)+H(1s; $^2S$) | $B^2\Sigma^+$ | [44] | 1.56-9.5 | fOH5c | {fOH5c} | |
| | O($2s^22p^3(^4S°)3s$; $^3S°$) +H(1s; $^2S$) | $^2\Sigma$ | | — | | {fOH6, 2xfOH7} | |
| | | $^4\Sigma$ | | — | | | |
| | O($2s^22p^3(^4S°)3s$; $^5S°$)+H(1s; $^2S$) | $^4\Sigma$ | | — | | {2xfOH6, 3xfOH7} | |
| | | $^6\Sigma$ | | — | | | |
| | O$^+$+H$^-$ | | [41] | 2.3-10 | fOH17 | {fOH17} | used for other excited levels, too |
| | O$^-$($^2P$)+H$^+$ | $3^2\Pi$ | [44] | 1.54-9.4 | fOH18a | {2xfOH18a, fOH18b} | |
| | | $C^2\Sigma^+$ | | 1.54-9.4 | fOH18b | | |
| OH$^+$ | O$^+$($2s^22p^3$; $^4S°$)+H(1s; $^2S$) | $X^3\Sigma^-$ | [41] | 0.13 -10 | fOH6 | {3xfOH6, 5xfOH7} | |
| | | $^5\Sigma^-$ | [45] | 1.1-20 | fOH7 | | |
| | O($2s^22p^4$; $^3P_0$)+H$^+$ | $A^3\Pi$ | [46] | 1.3 - 8 | fOH8 | {fOH8} | |
| | O($2s^22p^4$; $^3P_2$)+H$^+$ | $A^3\Pi$ | [46] | 1.3 - 8 | fOH8 | | |
| | O($2s^22p^4$; $^3P_1$)+ H$^+$ | $^3\Sigma^-$ | [45] | 1.14-20 | fOH9 | {fOH9} | Onvlee associates with O$^+$($^4S$)+H($^2S$) |
| | O($2s^22p^4$; $^1D$)+ H$^+$ | $a^1\Delta$ | | 1.3 - 8 | fOH10 | {2xfOH10, fOH11, 2xfOH12} | |
| | | $b^1\Sigma$ | [46] | 1.3 - 8 | fOH11 | | |
| | | $c^1\Pi$ | | 1.3 - 8 | fOH12 | | |
| | O($2s^22p^4$; $^1S$)+H$^+$ | $2^1\Sigma^+$ | [45] | 1-20 | fOH20 | {fOH20} | |
| | O($2s^22p^3(^4S°)3s$; $^3S°$) +H$^+$ | $^3\Sigma$ | | — | | {fOH6} | |
| | O($2s^22p^3(^4S°)3s$; $^5S°$)+H$^+$ | $^5\Sigma^-$ | | — | | {fOH7} | |
| | O$^+$($2s^22p^3$; $^2D°$)+H(1s; $^2S$) | $2^3\Pi$ | | 1.12-20 | fOH21 | {8xfOH21, 6xfOH22, fOH23, 3xfOH24, 2xfOH25} | |
| | | $2^1\Pi$ | | 1.12-20 | fOH21 | | |
| | | $^3\Delta$ | | 1.06-20 | fOH22 | | |
| | | $^1\Sigma^-$ | | 1.12-20 | fOH23 | | |
| | | $3^3\Sigma^-$ | | 1.08-20 | fOH24 | | |
| | | $2^1\Delta$ | [45] | 1.16-20 | fOH25 | | |
| | O+($2s^22p^3$; $2P°$)+H(1s; $^2S$) | $3^1\Pi$ | | 1.16-20 | fOH26 | {2xfOH26, fOH27, 3xfOH28, 6xfOH29} | |
| | | $3^1\Sigma^+$ | | 1.32-20 | fOH27 | | |
| | | $^3\Sigma^+$ | | 1.12-20 | fOH28 | | |
| | | $3^3\Pi$ | | 1.16-20 | fOH29 | | |
| OH$^{2+}$ | O$^{2+}$($^3P$)+H($^2S$) | $^4\Sigma^-$ | | 2.3-12 | fOH30 | {fOH30,2xfOH31} | |
| | | $^4\Pi$ | [47] | 2.14-12 | fOH31 | | |
| | O$^+$($2s2p^4$; $^4P$)+H$^+$ | $^4\Sigma^-$ | | 1.85-12 | fOH32 | {fOH32,2xfOH33} | used for other excitated levels, too |
| | | $^4\Pi$ | | 1.59-12 | fOH33 | | |
| OH$^-$ | O$^-$($^2P$)+H(1s; $^2S$) | $^1\Sigma^+$ | | 0.97 - 12.45 | fOH13 | {fOH13, 6xfOH14, 3xfOH15} | |
| | | $^3\Pi$ | [48] | 0.97 - 12.45 | fOH14 | | |
| | | $^3\Sigma$ | | 2.28 - 12.45 | fOH15 | | |
| | | $^1\Pi$ | | — | | | |
| | O$^-$($^2P$)+H(2p; $^2P_{1/2}$) | $^1\Delta$ | [48] | 0.97 - 12 | fOH16 | {fOH16} | |
| | | all other | | — | | | |
| | O+H$^-$ | | [41] | 0.2-10 | fOH19 | {fOH19} | used for other excited levels, too |
| ArO | Ar+O | | [41] | 0.21-12 | fArO1 | {fArO1} | used for other excited levels, too |
| | Ar$^+$+O$^-$ | | [41] | 1.7-10 | fArO3 | {fArO3} | used for other excited levels, too |
| ArO$^+$ | Ar$^+$($^2P$)+O($^3P$) | $1^2\Pi$ | [41] | 0.3-10 | fArO2 | {4xfArO2, fArO10, 2fArO11,2xArO12, 4xArO13, 4xArO14, 6xArO15, 4xArO16} | |
| | | $^2\Sigma^+$ | | 2.83-18.89 | fArO10 | | |
| | | $^2\Delta$ | | 2.83-18.89 | fArO11 | | |
| | | $^2\Sigma^-$ | | 3.02-18.89 | fArO12 | | |
| | | $1^4\Pi$ | [49] | 3.02-18.89 | fArO13 | | |
| | | $2^4\Pi$ | | 3.02-18.89 | fArO14 | | |
| | | $^4\Sigma^+$ | | 3.02-18.89 | fArO15 | | |
| | | $^4\Delta$ | | 3.02-18.89 | fArO16 | | |
| | Ar+O$^+$ | $^4\Sigma$ | | 2.83-18.89 | fArO9 | {fArO9} | |
| ArO$^{2+}$ | Ar$^+$($^2P$)+O$^+$($^4S$) | $X^3\Sigma^-$ | | 1.5-4.6 | fArO4 | {fArO4, 2xfArO5} | |
| | | $^3\Pi$ | | 1.5-4.6 | fArO5 | | |
| | Ar$^+$($^2P$)+O$^+$($^2D$) | $^1\Delta$ | [50] | 1.5-4.6 | fArO6 | {fArO7, 2xfArO6, 2fArO8} | |
| | | $^1\Sigma^+$ | | 1.5-4.6 | fArO7 | | |
| | | $^1\Pi$ | | 1.5-4.6 | fArO8 | | |
| ArH | Ar$^+$+H$^-$ | | [51] | 2.8-11 | fArH16 | {fArH16} | |
| | Ar($^1S$)+H(1s $^2S$) | $X^2\Sigma^+$ | [41] | 0.1-10 | fArH2 | {fArH2} | |
| | Ar+H(excited states) | | [52] | | fArH1 | {fArH1} | all similar to the Ar($^1S$)+H$^+$ curve |
| ArH$^+$ | Ar($^1S$)+H$^+$ | $X^1\Sigma$ | | 0.5-12 | fArH1 | {fArH1} | |
| | Ar$^+$($3s^23p^5$; $^2P°_{3/2}$) +H(1s; $^2S$) | $b^3\Pi$ | | 0.4-12 | fArH7 | {3xfArH7,fArH8} | used for other excitatied levels of H, too |
| | | $B^1\Pi$ | | 0.5-12 | fArH8 | | |
| | Ar$^+$($3s^23p^5$; $^2P°_{1/2}$) +H(1s; $^2S$) | $a^3\Sigma$ | | 0.4-12 | fArH9 | {3xfArH9,fArH10} | |
| | | $A^1\Sigma$ | [53] | 0.5-12 | fArH10 | | |
| | Ar($^3P$)+H$^+$ | $c^3\Sigma$ | | 0.4-12 | fArH11 | {3xfArH11, 6xfArH12, 2xfArH13,fArH14} | used for other excited levels, too |
| | | $d^3\Pi$ | | 0.4-12 | fArH12 | | |
| | Ar($^1P$)+H$^+$ | $D^1\Pi$ | | 0.5-12 | fArH13 | | |
| | | $C^1\Sigma$ | | 0.5-12 | fArH14 | | |
| | Ar$^{2+}$($^3P$)+H$^-$($^1S$) | ? | [54] | 0.5-51 | fArH6 | {fArH6, fArH6a} | |
| | | ? | | 0.5-51 | fArH6a | | |
| ArH$^{2+}$ | Ar$^+$($3s^23p^5$; $^2P°_{1/2}$)+H$^+$ | $^2\Sigma$ | [55] | 2-18.8 | fArH3 | {fArH3} | |
| | Ar+($3s^23p^5$; $^2P°_{3/2}$)+H$^+$ | $^2\Pi$ | | 1.87-18.87 | fArH4 | {fArH4} | |
| | Ar$^{3+}$+H$^-$ | | | — | | {fArH6} | |
| | Ar$^{2+}$+H | | | — | | {fArH10} | |
| ArH$^-$ | H$^-$+Ar | | [56] | 2.6-16 | fArH15 | {fArH15} | |

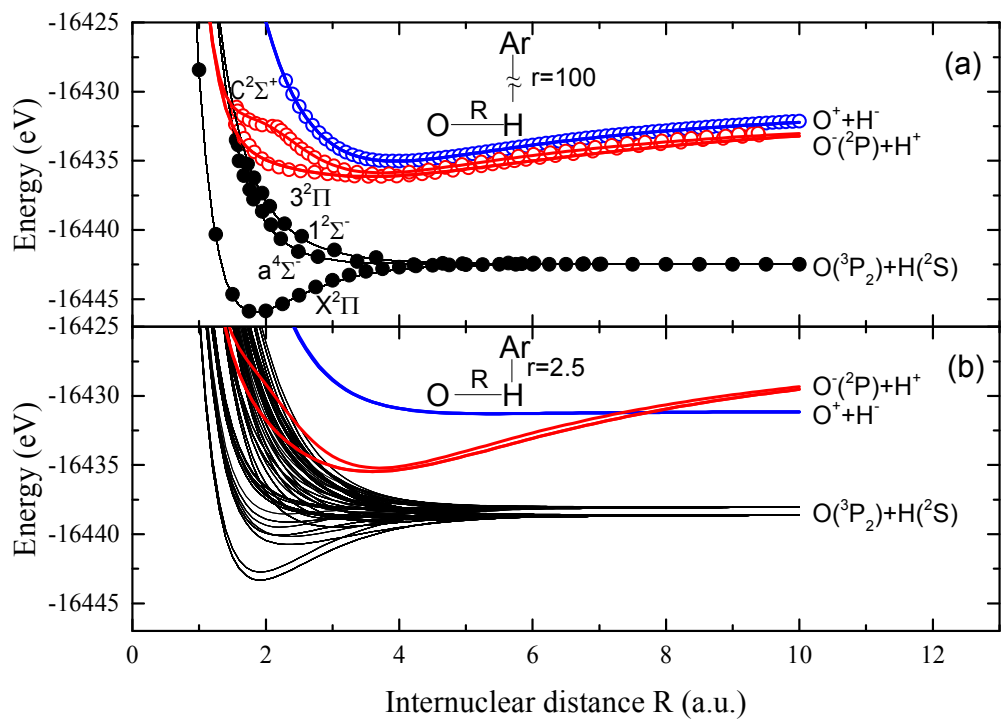

**Figure 13** (Color online) Energy levels of the Ar$^+$($^2$P) + OH system as function the OH distance. Dissociation limits for the OH radical and molecular terms are indicated in the figure. Lines are the resulting fit functions (see text). Black lines: atomic dissociation limit (O + H), red lines: ionic dissociation limits (O$^-$ + H$^+$), and blue lines: (O$^+$ + H$^-$). Circles: data from the literature see Table 3. Asymptotic levels are fitted to the sum of atomic energy levels from NIST database [40]. The (Ar, H, O) centers form a right triangle. The Ar–H distance r is 100 au. and 2.5 au. in panel (a) and (b) respectively. In panel (b), one can observe the splitting of the primary OH curves due to the nearness of Ar center.